\documentclass[conference]{IEEEtran}
\IEEEoverridecommandlockouts
\usepackage{cite}
\usepackage{amsmath,amssymb,amsfonts}
\usepackage{algorithmic}
\usepackage{graphicx}
\usepackage{textcomp}
\usepackage{xcolor}

\usepackage{booktabs} 

\usepackage[english]{babel}
\usepackage{moresize}
\usepackage{amsmath}

\usepackage{algorithmic} 
\usepackage{balance}
\usepackage{marvosym}
\usepackage{comment}
\usepackage{paralist}
\usepackage{multirow}

\usepackage{filecontents}

\usepackage[english]{babel}
\usepackage[latin1]{inputenc}
\usepackage{mathrsfs}
\usepackage{graphicx}
\usepackage{amssymb}
\usepackage{url}
\usepackage{subfigure}
\usepackage{amsmath}
\usepackage{enumitem}
\usepackage[linesnumbered,algoruled,boxed,lined]{algorithm2e}
\usepackage{adjustbox}
\usepackage{amssymb}
\usepackage{times}
\usepackage{hyperref}

\usepackage{pgfplots}
\usetikzlibrary{pgfplots.dateplot}

\usepackage{filecontents}
\definecolor{tblue}{RGB}{31,119,180}
\definecolor{torange}{RGB}{255,127,14}
\definecolor{tgreen}{RGB}{44,160,44}
\definecolor{tred}{RGB}{214,39,40}
\definecolor{tpurple}{RGB}{148,103,189}

\usepackage{filecontents}
\usepackage{balance}

\newcommand{\hide}[1]{} 

\newcommand{\etal}{\textit{et al}.}

\newcommand{\ie}{\textit{i}.\textit{e}.}
\newcommand{\eg}{\textit{e}.\textit{g}.} 
\newcommand{\wrt}{\textit{w}.\textit{r}.\textit{t}}

\def\BibTeX{{\rm B\kern-.05em{\sc i\kern-.025em b}\kern-.08em
    T\kern-.1667em\lower.7ex\hbox{E}\kern-.125emX}}
    
\begin{document}



\title{Disentangled Graph Social Recommendation

\noindent \thanks{\textbf{\dag Equal contribution. *Corresponding author: Chao Huang.}}}

\def\model{DGNN}
\def\full{Disentangled Heterogeneous Graph Neural Network}



\author{\IEEEauthorblockN{Lianghao Xia$^{1,\dag}$, Yizhen Shao$^{2,\dag}$, Chao Huang$^{1,*}$, Yong Xu$^2$, Huance Xu$^2$, Jian Pei$^3$}
\IEEEauthorblockA{University of Hong Kong$^1$, South China University of Technology$^2$, Duke University$^3$ \\
aka\_xia@foxmail.com, chaohuang75@gmail.com, \{csyzshao, cshuance.xu\}@mail.scut.edu.cn} yxu@scut.edu.cn, j.pei@duke.edu}



\maketitle

\begin{abstract}
Social recommender systems have drawn a lot of attention in many online web services, because of the incorporation of social information between users in improving recommendation results. Despite the significant progress made by existing solutions, we argue that current methods fall short in two limitations: (1) Existing social-aware recommendation models only consider collaborative similarity between items, how to incorporate item-wise semantic relatedness is less explored in current recommendation paradigms. (2) Current social recommender systems neglect the entanglement of the latent factors over heterogeneous relations (\eg, social connections, user-item interactions). Learning the disentangled representations with relation heterogeneity poses great challenge for social recommendation. In this work, we design a \underline{D}isentangled \underline{G}raph \underline{N}eural \underline{N}etwork (\model) with the integration of latent memory units, which empowers \model\ to maintain factorized representations for heterogeneous types of user and item connections. Additionally, we devise new memory-augmented message propagation and aggregation schemes under the graph neural architecture, allowing us to recursively distill semantic relatedness into the representations of users and items in a fully automatic manner. Extensive experiments on three benchmark datasets verify the effectiveness of our model by achieving great improvement over state-of-the-art recommendation techniques. The source code is publicly available at: \url{https://github.com/HKUDS/DGNN}.

\end{abstract}


\section{Introduction}
\label{sec:intro}



Recommender systems which aim to suggest items with the learning of user's personalized interests, have provided essential web services (\eg, E-commerce sites~\cite{li2020hierarchical}, online review systems~\cite{lyu2021reliable} and advertising platforms~\cite{liu2021neural}) to alleviate the information overload problem~\cite{xie2022contrastive}. To address the sparse data limitation of conventional collaborative filtering models, there exist many recommendation paradigms leveraging social relationships between users, to enhance the user-item interaction modeling with external information source. These approaches explicitly characterize the cross-user influence with respect to their interaction preference in recommendation~\cite{ji2021you}.

Motivated by the prevalence of graph neural networks (GNNs), recently emerged social recommendation methods utilize graph neural encoders to shine a light on modeling graph structure of social connections, and iteratively aggregate feature information from local neighborhoods. For example, recent efforts (\eg, DiffNet~\cite{wu2019neural}, MHCN~\cite{yu2021self} and KCGN~\cite{2021knowledge}) employ graph convolution to capture the social-aware collaborative filtering signals and guide the user representation learning. GraphRec~\cite{fan2019graph} and DANSER~\cite{wu2019dual} are developed based on graph attention mechanism to discriminate relations that connect interacted users. Such GNN-based social-aware recommender systems have generated state-of-the-art performance by modeling the high-order connectivity among users and items. Additionally, another relevant attempts for jointly exploiting user-user and user-item relational structure lie in the feature transfer learning from social domain to user-item interaction encoding process~\cite{chen2019efficient,xiao2017learning}.

Despite their effectiveness, we argue that existing social recommender systems fall short in two limitations:

(1) The rich semantic relatedness among items remains unexplored by most existing learning solutions. In real-world recommendation scenarios, there typically exist dependencies across items, \eg, product categories/functionality, spatial similarities of venues~\cite{tanjim2020attentive,hu2018leveraging}. Such rich semantic relatedness among items can help explore their latent dependencies, which is helpful to understand complex interests of users~\cite{wang2019kgat,xin2019relational}. As a result, user's preference over different items may not only be affected by his/her social connections, but also be inferred from the fine-grained relational knowledge on items. Despite the above benefits, incorporating the cross-item dependencies in social recommendation is challenging due to the heterogeneity nature of various relations. \\\vspace{-0.12in}


(2) Most of current social recommender systems ignore the fact that connections are driven by complex factors. For instance, user's intent on interacting (\eg, click, or purchase) an item may be influenced by diverse factors due to different item characteristics, such as the brand and color of products, director of a movie~\cite{wang2020disentangled,wang2022disentangled}. The overlook of finer-grained user interest with the factor-level representation learning, may produce suboptimal recommendation results. Moreover, in real-life social recommendation scenario, users are socially connected due to multifaceted motives~\cite{liu2019single,epasto2019single}, \eg, communities with disparate interests, colleagues, or family members. If we represent user-wise influence without the disentanglement of such social polysemy, the learned user preference is hard to be reflective of multiple social contexts. Therefore, the heterogeneous relations driven by complex latent factors, brings an urge for the model to encode factorized embeddings pertinent to type-specific relation semantics.

State-of-the-art recommender systems are proposed to add disentangled representation learning into the user-item interaction modeling. However, they merely focus on single type of relation disentanglement, which are limited to incorporate heterogeneous relational semantics into recommender systems. In other words, an important fact in recommendation has been ignored: diverse semantics with relation heterogeneity can be utilized to enhance the user preference learning in recommender system. Hence, we need to learn disentangled user/item factorized embeddings with the awareness of heterogeneous side context for enhancing the representation power of neural recommendation models in real-life applications.

One feasible way of modeling heterogeneous relations in social recommendation is to rely on the heterogeneous graph learning approaches~~\cite{shi2018heterogeneous,fan2019metapath,wang2019heterogeneous}. Specifically, those methods exploit the connection structures of user-user and item-item relations into the graph learning model. However, these methods heavily rely on manually designed meta-paths between users/items with the requirement of specific domain knowledge, which can hardly be adaptive in diverse recommendation scenarios. They either simply keep distinct transformation weights during the feature representation for either node type or edge type alone. It makes them insufficient to comprehensively capture heterogeneous relational context from both user and item side, as well as the underlying interaction context between user- and item-wise relations. 

While having realized the vital role of encoding disentangled relation heterogeneity in recommendation, it is a non-trivial task due to the following key challenges: i) Learning the disentangled factors with relation heterogeneity brings the challenge of graph neural networks. In our method, we develop a node- and edge-type dependent memory-augmented network to preserve dedicated semantic representations for different types of interactions, \ie, user-user, item-item and user-item relationships. ii) Capturing the implicit inter-dependencies among different encoded disentangled relation factors. The new designed graph neural network should enable the recommender system to learn the cross-factor inter-dependencies for expressive disentangled representation learning.


In light of these limitations and challenges, we propose a \underline{\textbf{D}}isentangled \underline{\textbf{G}}raph \underline{\textbf{N}}eural \underline{\textbf{N}}etwork (\model), to study the social recommendation with the learning of disentangled heterogeneous factors. To handle relation heterogeneity with disentangled relation modeling, we develop a node- and edge-type dependent memory-augmented network to preserve  dedicated feature representations for different types of interactions, \ie, user-user, item-item and user-item relationships. Particularly, \model\ utilizes external memory units with differentiable embedding propagation operators, allowing graph neural architecture to explicitly capture the heterogeneous graph relations for social recommendation. Additionally, instead of parameterizing each type of relations, the introduced memory neural layers endow the relation heterogeneity encoding under disentangled latent representation spaces in a fully automatic and interact manner, without customized meta paths. 


To summarize, we make the following contributions:

\begin{itemize}[leftmargin=*]

\item We emphasize the importance of integrating the heterogeneous relationships with latent factor disentanglement in social recommender systems. It empowers the user preference representation paradigm with the exploration of attractive source of information from both user and item domains.

\item We propose a disentangled graph neural networks \model\ which generalizes the relation heterogeneity encoding by maintaining the relation-aware disentangled representations. Our proposed \model\ is built upon the heterogeneous graph memory-augmented message passing.

\item Extensive experiments on three real-world datasets demonstrate that \model\ significantly beats various state-of-the-art recommendation methods. In addition, the elaborated model ablation study helps justify the model effectiveness and component-wise impact in performance improvement.

\end{itemize}


\section{Related Work}
\label{sec:relate}

\subsection{Social-aware Recommendation Methods}
To improve the recommendation results, many social recommendation methods have been proposed to incorporate the online social relationships between users into the recommendation framework as side information~\cite{liu2019social,jin2020partial,liu2018social}. Most traditional methods (\eg, Sorec~\cite{ma2008sorec}, TrustMF~\cite{yang2016social}) are built based on the matrix factorization architecture to project users into latent factors. The common rationale behind those approaches is that users are more likely to share similar interests over items with their socially connected friends~\cite{tang2013social}.

Deep learning-based social recommendation models have received increasing attention, due to the ability of neural networks for knowledge representation~\cite{yu2020enhance,fu2021dual}. Specifically, some studies focus on applying graph convolutional network to simultaneously model the user-user and user-item relationships, like DiffNet~\cite{wu2019neural}, RecoGCN~\cite{xu2019relation} and KCGN~\cite{2021knowledge}. Additionally, attention mechanisms have been introduced to differentiate influence among users for characterizing their preference, such as SAMN~\cite{chen2019social} and GraphRec~\cite{fan2019graph}. For example, GraphRec distinguishes the strength of social ties when aggregating information from both social and user-item interaction graph. Motivated by self-supervised learning, data augmentation is applied in recent social recommender systems MHCN~\cite{yu2021self} and SMIN~\cite{long2021social}. However, most of those recommendation methods disregard the latent factors underlying heterogeneous relationships. To fill this gap, this work therefore seeks for a new social recommender system that integrates the disentangled representation learning with heterogeneous semantics under a graph neural architecture.\\\vspace{-0.1in}


\subsection{Graph Neural Network for Recommendation}
Due to the strength in representation learning over graph-structured data, a line of research for recommendation focuses on enhancing the user-item interaction modeling with graph neural architectures~\cite{he2020lightgcn,chen2020revisiting,xia2022hypergraph}. Inspired by the effectiveness of spectral graph convolutional network, NGCF~\cite{wang2019neural} proposes to capture high-order relationship between user and item by performing the convolutional operations. Furthermore, another line of GNN-based recommender systems explores the spatial GNNs through attentively aggregating information from neighboring nodes, such as KGAT~\cite{wang2019kgat} and DGRec~\cite{song2019session}.

In addition, amongst the GNN research, heterogeneous graph representation has become the promising solution to integrate the diverse relational context into the node representation~\cite{wang2019heterogeneous,fu2020magnn,chen2023heterogeneous,hu2020heterogeneous,wei2022contrastive}. For example, HERec~\cite{shi2018heterogeneous} attempts to incorporate various side information to enhance the user preference learning based on the generated meta-path connections. Multi-typed user-item interactions (\eg, click, purchase) are considered to encode relation heterogeneity~\cite{gu2022hybridgnn,xie2021sequential}. However, the accurate prediction results from most of them largely rely on the effectiveness of the constructed meta-path-based connections, which requires domain-specific knowledge. Different from them, our \model\ model automatically capture heterogeneous relationships across users and items, by designing the memory-augmented message passing scheme. Additionally, the latent factor encoding with disentangled representations has largely been unexplored in existing recommenders which consider heterogeneous context.

\subsection{Disentangled Learning for Recommendation}
There exist some recent studies focusing on learning disentangled representations from user-item interactions to distill the latent factors driving the observed connections~\cite{qian2022intent}. For example, DGCF~\cite{wang2020disentangled} designs routing mechanism with capsule network to model the disentangled relationships between users and items. Chen~\etal~\cite{chen2021curriculum} propose a curriculum learning-based method to disentangle multi-typed feedback of users. In multimedia recommendation domain, the multi-modal features are incorporated into the disentangled representation learning in a weakly supervised way~\cite{wang2021multimodal}. 

While there exist some works on disentangled learning for recommendation, our new recommender system differs from those studies from the following two aspects: i) most of those models are designed to model the homogeneous relations in recommender system, which cannot be easily adaptive to disentangle the diverse factors behind the heterogeneous relationships due to their various semantics. To fill this gap, our disentangled graph neural network is designed to maintain customized factorized representations for relation heterogeneity and distill the relational knowledge in a fully automatic manner. ii) The complex dependencies among the encoded latent factors are ignored in most current studies. In contrast, our \model\ method captures the latent factor-wise inter-dependencies with our designed differentiable memory networks under multiple latent representation spaces. By doing so, the relation-aware latent factors can effectively preserve the disentangled heterogeneous semantics.




\subsection{Context-aware Recommender Systems}
There exist some relevant research works focusing in developing context-aware recommender systems with the consideration of various context in different recommendation scenarios~\cite{adomavicius2022context, kulkarni2020context}.
From the user dimension, STARS~\cite{kulkarni2020context} combines the user-wise relations from online social network and user-specific contextual features to enhance collaborative filtering. From the item content dimension, LBSNs~\cite{aliannejadi2018personalized} learns the mapping from items' contextual features to users' preference tags in Point-of-Interest recommendation. CARL~\cite{wu2019context} proposes to fuse the textual context information of items and the interaction-based embeddings for better representation learning. Additionally, knowledge graphs have also been considered as useful contextual signals to be incorporated into recommender system to improve performance in knowledge-aware recommenders KGIN~\cite{wang2021learning}, CASR~\cite{mezni2021context} and KGCL~\cite{yang2022knowledge}. 

\section{Preliminaries}
\label{sec:model}

\begin{figure*}[t]
	\centering
	\includegraphics[width=0.98\textwidth]{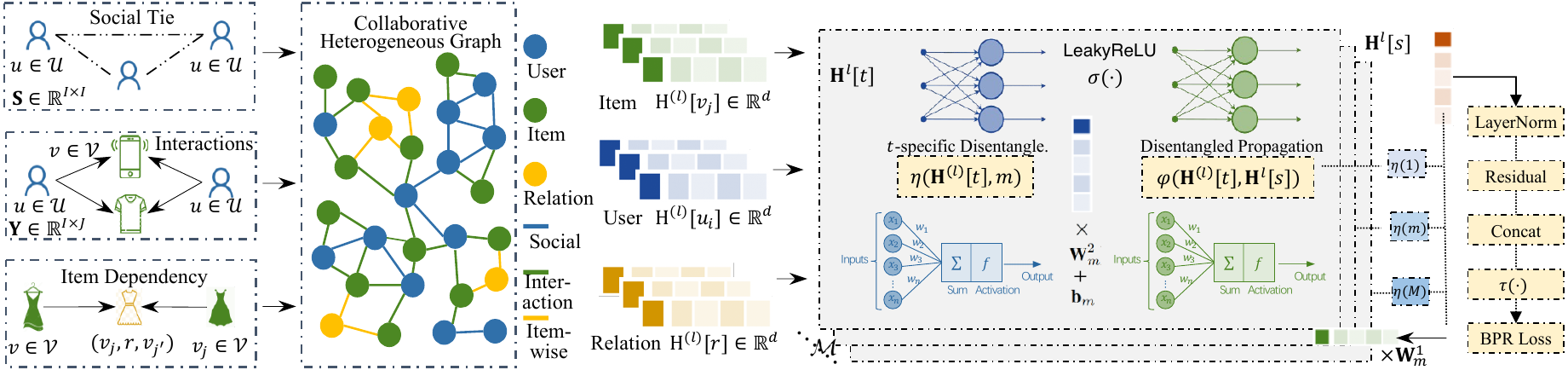}
	\caption{The model flow of the proposed \model\ architecture. The collaborative heterogeneous graph together with the initial user/item/relation-node embeddings are fed into our \model\ composed of the disentangled heterogeneity encoder and the heterogeneous graph message aggregator. $\eta(m)$ represents the learned importance weight of the $m$-th memory unit corresponding to the encoded factorized embeddings.}
	\label{fig:framework}
 	\vspace{-0.1in}
\end{figure*}

We consider a scenario with the sets of users $\mathcal{U} = \{u_1,...,u_i,...,u_I\}$ and items $\mathcal{V} = \{v_1,...,v_j,...,v_J\}$. The user-item interactions are denoted by matrix $\textbf{Y} \in \mathbb{R}^{I\times J} = \{ y_{i,j} | u\in \mathcal{U}, v\in \mathcal{V} \}$, where $y_{i,j}=1$ if the interaction (\eg, browse, or purchase) between user $u_i$ and item $v_j$ is observed and $y_{i,j}=0$ otherwise. In addition to the interaction matrix $\textbf{Y}$, we also have the social connections between users which are represented by the defined user social matrix $\textbf{S} \in \mathbb{R}^{I\times I}$, where each entry $s_{i,i'}=1$ if there exists a social tie between user $u_i$ and $u_{i'}$ and zero otherwise. In this work, we propose to enhance the social recommendation with the incorporation of item relations. Hence, given an item pair ($v_j$, $v_{j'}$), their relationships are defined as an entity-relation-entity triple $(v_j, r, v_{j'})$, where $r\in \mathcal{R}$ is the intermediate relation node for cross-item meta relation (\eg, categorical relations between items), where $v_j$, $v_{j'}\in \mathcal{V}$. We use the item-relation connections $(v_j, r)$ to construct the item relation matrix $\textbf{T}\in\mathbb{R}^{J\times R}$, where $R$ denotes the number of relations. \\\vspace{-0.12in}


\noindent \textbf{Task Description}. With the above definitions, we define the task of social recommendation with item relations as: \textbf{Input} the user-item interaction history records $\textbf{Y}$, the user social relation matrix $\textbf{S}$, and the item-relation matrix $\textbf{T}$. \textbf{Output} a trained model $\xi(\cdot)$ that forecasts the preference of each user $u_i$ over unobserved items $v_j$ by $\hat{y}_{i,j} = \xi(u_i, v_j; \textbf{Y}, \textbf{S}, \textbf{T})$
\section{Methodology}
\label{sec:solution}


In this section, we present the details of our \model\ framework. Our new method consists of three key components: i) Memory-augmented relation heterogeneity encoder which parameterizes the relation heterogeneity into vertex- and edge-type dependent embeddings. ii) Disentangled message aggregation with relation heterogeneity that fuses the disentangled relational context from both interactive patterns and side information. iii) Model forecasting stage which incorporates the heterogeneous factorized embeddings into the model optimized objective for recommendation. The overall architecture of our proposed \model\ is shown in Figure~\ref{fig:framework}.



\subsection{Collaborative Heterogeneous Graph}
To jointly preserve the user-item interactions $\textbf{Y}$, the user-user social connections $\textbf{S}$, and the item-wise relations $\textbf{T}$, we define a unified graph structure $\mathcal{G}=(\mathcal{D}, \mathcal{E}, \mathcal{A}, \mathcal{B})$ where each vertex $d \in \mathcal{D}$ and each edge $e \in \mathcal{E}$ are associated with their mapping functions: $\mathcal{D} \rightarrow \mathcal{A}$ and edges $\mathcal{E} \rightarrow \mathcal{B}$. Here, $\mathcal{A}$ and $\mathcal{B}$ denotes the sets of vertices and relation edge types, respectively. In graph $\mathcal{G}$, we characterize the relation heterogeneity with various connections across users and items by performing the integration as follows:
\begin{align}
    \mathcal{D} = \mathcal{U} \cup \mathcal{V}\cup \mathcal{R}; ~~~\mathcal{E} = \mathcal{S} \cup \mathcal{T} \cup \mathcal{Y}
\end{align}
where $\mathcal{S}, \mathcal{T}, \mathcal{Y}$ denote the sets of observed edges in the adjacent matrices $\textbf{S}, \textbf{T}, \textbf{Y}$, respectively (\eg~$\mathcal{Y}=\{(u_i, v_j):y_{i,j}=1\}$). With the unified heterogeneous graph integrating three types of relations, we then design our heterogeneous GNN architecture with disentangled factor representations.

\subsection{Disentangled Heterogeneous Graph Memory Network}

Despite the progress, most existing methods for heterogeneous graph data barely pay attention to the latent factors that generate the complex semantics of the heterogeneous data. For example, HGT~\cite{wang2019heterogeneous} assigns each node/edge type with an individual parameter set directly. This hinders the deep understanding of heterogeneous data such as latent type-wise dependencies. Inspired by the strength of memory neural networks in disentangled representation learning~\cite{santoro2016meta, chen2019social}, our \model\ proposes to augment the graph neural network with differentiable memory components under multiple latent representation space. \model\ adaptively learns the connections between node/edge types and latent factors, so as to better preserve the disentangled heterogeneous semantics.



Our \model\ takes $\mathcal{G}$ as the input computation graph for information propagation. During the message passing, we first perform the local feature transformation and nonlinear activation, and then aggregate relation-aware contextual representations. Formally, it can be represented with the following form (from the $(l)$-th layer to $(l+1)$-th layer):
\begin{align}
\label{eq:general}
\textbf{H}^{(l+1)}[t] \leftarrow \mathop{\text{Aggre}}\limits_{\forall s\in \mathcal{N}(t)} \Big ( \varphi( \textbf{H}^{(l)}[t], \textbf{H}^{(l)}[s], e_{s,t}) \Big )
\end{align}
\noindent where we define $\textbf{H}^l[t]$ as the latent representation of target node $t$ for the $l$-th graph layer. $\mathcal{N}(t)$ denotes the set of neighboring nodes (\ie, source node $s$) of node $t$. Edge $e_{s,t}$ connects node $s$ and $t$. In the message passing paradigm, \model\ consists of two key operators: relation heterogeneity encoder $\varphi(\cdot)$ and embedding aggregation function \text{Aggre}$(\cdot)$.

\subsubsection{Memory-Augmented Relation Heterogeneity Encoder}

\begin{figure}
    \centering
    \includegraphics[width=0.95\columnwidth]{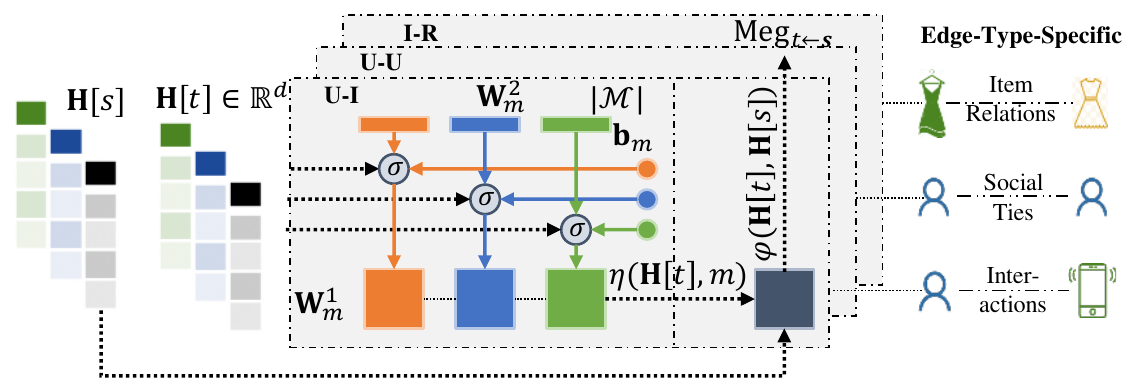}
    \caption{Illustration for the message passing based on our disentangled heterogeneous graph encoder for different user and item relations.}
    \label{fig:memoatt}
\end{figure}

To capture the heterogeneous characteristic in the knowledge-enhanced social recommendation, we propose to parameterize the relation heterogeneity into vertex- and edge-type dependent embedding projection through external memory units (as shown in Figure~\ref{fig:memoatt}. Specifically, we first define $\mathcal{M}$ be the set of memory units corresponding to factor prototype learning $\varphi(\cdot)$ for type-specific relation semantics as below:
\begin{align}
\label{eq:memoatt}
\varphi( \textbf{H}^{(l)}[t], \textbf{H}^{(l)}[s]) &= \Big ( \sum_{m=1}^{|\mathcal{M}|} \eta(\textbf{H}^{(l)}[t], m) \textbf{W}_{m}^1 \Big ) \textbf{H}^{(l)}[s] \nonumber \\
\eta(\textbf{H}^{(l)}[t], m) &= \sigma(\textbf{H}^{(l)}[t] \cdot \textbf{W}_{m}^2 + \textbf{b}_{m})
\end{align}
\noindent where $\eta(\cdot)$ represents the target node-specific embedding function. The trainable transformation matrices and bias terms are denoted as: $\textbf{W}_{m}^1 \in \mathbb{R}^{d\times d}$, $\textbf{W}_{m}^2 \in \mathbb{R}^{d}$ and $\textbf{b}_{m} \in \mathbb{R}$. The encoded feature embeddings of source and target nodes are represented as $\textbf{H}^{(l)}[s] \in \mathbb{R}^{d}$ and $\textbf{H}^{(l)}[t] \in \mathbb{R}^{d}$, respectively. We apply the LeakyReLU as the activation function $\sigma(\cdot)$, \ie, $\sigma(x) = \text{max}(x, \alpha x)$. The negative slope $\alpha$ is set as 0.2 for better gradient back-propagation. Our memory-augmented network allows the graph encoder to learn relation representation with hierarchical non-linear property. To maintain dedicated representations for different types of nodes (\eg, users, item) and edges (\ie, interactions, social connections, inter-item relationships), we also perform the encoding for edge type-specific relation with non-sharing hyperparameter space. By doing so, the learned disentangled representations are able to preserve diverse latent factors pertinent to different relations in the collaborative heterogeneous graph $\mathcal{G}$.

\subsubsection{Message Aggregation with Relation Heterogeneity}
\begin{figure}
	\centering
	\includegraphics[width=0.48\textwidth]{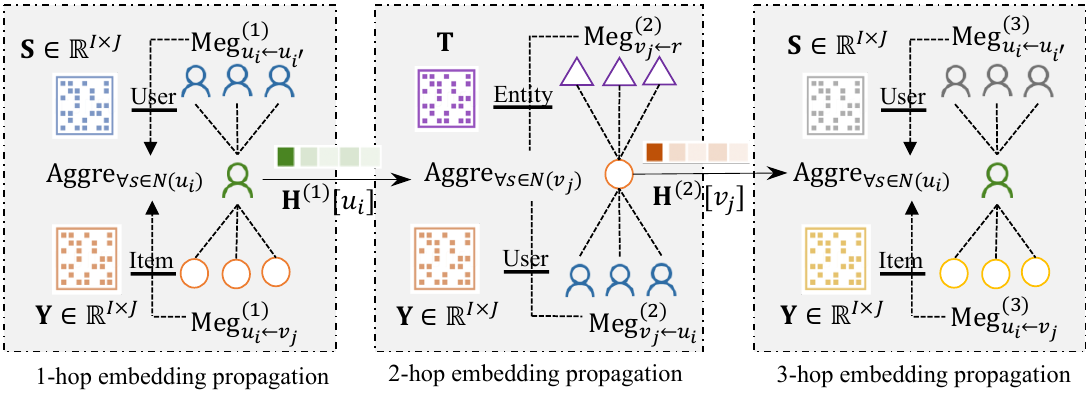}
	\caption{Illustration of message passing with disentangled relation heterogeneity across users and items under three graph layers in our model.}
	\label{fig:message_passing}
\end{figure}

After encoding the heterogeneous relation properties from local neighbor interactions with $\textbf{H}^{(l)}[v]$, we next aggregate message from different information sources from both user and item domains (illustrated in Figure~\ref{fig:message_passing}). For example, we fuse the relational context from both interactive patterns and social influence in the message aggregator for users ($u_i$) as:
\begin{align}
    \label{eq:userAgg}
    \textbf{H}^{(l+1)}[u_i] &= \frac{1}{|\mathcal{N}_{u_i}^S|+|\mathcal{N}_{u_i}^Y|} \Big ( \sum_{u_i' \in \mathcal{N}_{u_i}^S} \varphi(\textbf{H}^{(l)}[u_i'], \textbf{H}^{(l)}[u_i])  \nonumber\\
 + & \sum_{v_j \in \mathcal{N}_{u_i}^Y} ( \sum_{m=1}^{|\mathcal{M}|} \eta(\textbf{H}^{(l)}[v_j], m) \textbf{W}^{m,1}_{i\leftarrow j} \Big ) \textbf{H}^{(l)}[u_i] \Big )
\end{align}
\noindent where $|\mathcal{N}_{u_i}^S|$ and $|\mathcal{N}_{u_i}^Y|$ denote the number of neighboring nodes of $u_i$, in the user-user social graph and in the user-item interaction graph, respectively. $\textbf{W}_{i\leftarrow j}^{m,1}\in\mathbb{R}^{d\times d}$ denotes the transformation matrix for mapping from the item representation space to the user representation space, with respect to the $m$-th memory unit of disentangled factor.

With the incorporation of knowledge-aware item relations, the embedding propagation process for item side can be formally presented as follows:
\begin{align}
\label{eq:itemAgg}
\textbf{H}^{(l+1)}[v_j]= \rho_{i,j} ( \sum_{v_j \in \mathcal{N}_{v_j}^Y} \text{Meg}^{(l)}_{v_j \leftarrow u_i} + \sum_{r \in N_{v_j}^T} \text{Meg}^{(l)}_{v_j \leftarrow r} ) 
\end{align}




\noindent where $\rho_{i,j}$ indicates the normalization term as $\rho_{i,j}=1/({|N_{v_j}^Y|+|N_{v_j}^T|})$. The propagated message ($\text{Meg}_{v_j \leftarrow u_i}$ and $\text{Meg}_{v_j \leftarrow r}$) is determined by the memory-based encoding function $\varphi(\cdot)$. Furthermore, the embedding propagation between items (\eg, $v_j$) and meta relation node ($r$) is shown below:
\begin{align}
\label{eq:relAgg}
\textbf{H}^{(l+1)}[r] = \frac{1}{|\mathcal{N}_r|} \sum_{v_j\in N_r} ( \sum_{m=1}^{|\mathcal{M}|} \eta(\textbf{H}^{(l)}[v_j], m) \textbf{W}^{m,1}_{r \leftarrow j} ) \textbf{H}^{(l)}[r]
\end{align}
\noindent where $\mathcal{N}_r$ denotes the set of neighbors for the meta relation node $r$ in the graph structure. $\textbf{W}_{r\leftarrow j}^{m,1}\in\mathbb{R}^{d\times d}$ denotes the $m$-th memory-unit-specific transformation for mapping from the item space to the meta relation node space.

We further generalize the heterogeneous message aggregation with the incorporation of self-propagation and layer normalization~\cite{ba2016layer} to stabilize the network training:
\begin{align}
\label{eq:layerNorm}
\widetilde{\textbf{H}}^{(l+1)}[v] = \sigma( \omega_1 \odot \frac{\textbf{H}^{(l+1)}[v]- \mu}{\sqrt{\sigma^2+\epsilon}}+ \omega_2 ) \\\nonumber
+ \varphi( \textbf{H}^{(l)}[v])
\end{align}
\noindent where $\omega_1$ and $\omega_2$ are learned scaling factors and bias terms. $\mu$ and $\sigma$ respectively denote the mean and variance of input vector $\textbf{H}^{(l+1)}[v]$. $\odot$ denotes the element-wise multiplication operator. For the self-loop, instead of directly adding the embeddings from the last GNN iteration, \model\ also applies the relation heterogeneity encoder $\phi(\cdot)$.
To make fully use of the multi-order node embeddings, we further perform the cross-layer ($L$) high-order embedding aggregation as follows:
\begin{align}
{\textbf{H}}^*[v] = \text{LayerNorm}( \widetilde{\textbf{H}}^{(0)}[v] \mathbin\Vert \widetilde{\textbf{H}}^{(1)}[v] \mathbin\Vert ... \mathbin\Vert \widetilde{\textbf{H}}^{(L)}[v])    
\end{align}
\noindent where $\textbf{H}^*[v] \in \mathbb{R}^{d}$ denotes the final node embeddings for vertex $v$. $\text{LayerNorm}(\cdot)$ denotes the layer normalization.

\subsection{Model Forecasting Phase}
To inject the social influence into our forecasting phase of \model, we refine the learned user embedding with the representation recalibration function $\tau(\cdot)$ shown as follows:
\begin{align}
    \tau(\textbf{H}^*[u_i]) = \frac{1}{|\mathcal{N}_{u_i}^S|+1}\Bigm(\sum\limits_{u_i'\in\mathcal{N}_{u_i}^S} \textbf{H}^*[u_i'] + \textbf{H}^*[u_i]\Bigm)
\end{align}
\noindent The function averages the socially-connected user embeddings to directly incorporate social information in the following prediction phase. We formally present it as follows:
\begin{align}
& \xi(u_i,v_j) = (\textbf{H}^*[u_i] + \tau(\textbf{H}^*[u_i]))^\top \cdot \textbf{H}^*[v_j] \nonumber\\
&= \textbf{H}^{*\top}[u_i] \cdot \textbf{H}^*[v_j] \\\nonumber 
&+  \frac{\textbf{H}^{*\top}[v_j]}{|N_{u_i}^S|+1} \Bigm(\sum_{u_i' \in N_{u_i}^S} \textbf{H}^*[u_i'] + \textbf{H}^*[u_i]\Bigm) 
\end{align}
\noindent\textbf{Optimization Objective}:
We define our optimization objective with the integrative loss of pairwise BPR loss and weight-decay regularization term in the following equation:
\begin{align}
\label{eq:loss}
\mathcal{L}=\sum_{(i,j^+,j^-)\in O} -\log ~\delta(\xi(i,j^+) - \xi(i,j^-)) + \lambda \mathbin\Vert \mathbf{\Theta} \mathbin\Vert^2 
\end{align}
\noindent The training data is denoted as $O=\{(i,j^+,j^-) | (i, j^+) \in \textbf{Y}^+, (u, j^-) \in \textbf{Y}^-\}$ ($\textbf{Y}^+$, $\textbf{Y}^-$ denotes the observed and unobserved interactions, respectively. Training parameters are denoted as $\mathbf{\Theta}$ and $\delta(\cdot)$ denotes the sigmoid activation function. The learning process of our \model\ is elaborated in Alg~\ref{alg:model}.

\subsection{Model Efficiency Analysis}
\subsubsection{Time Complexity Analysis}
To calculate the attention weights for each edges on the memory units, \model\ takes $O(|\mathcal{M}|\times |\mathcal{E}|\times d)$ complexity, where $|\mathcal{M}|$ denotes the number of memory units for each graph (\ie~the user-item collaborative graph, the user-user social graph, and the item-relation graph). With the attention scores, the time complexity to obtain the dedicated transformation for each edge (\ie~$\sum_{m=1}^{|\mathcal{M}|}\eta(\textbf{H}^{(l)}[t], m)\textbf{W}_m^1$) is $O(|\mathcal{M}|\times |\mathcal{E}|\times d^2)$. Then \model\ takes $O(|\mathcal{V}|\times d^2)$ time complexity to conduct embedding transformation, and needs $O(|\mathcal{E}|\times d)$ complexity for information propagation along the heterogeneous edges. Generally, $|\mathcal{V}|\times d < |\mathcal{E}|$ due to the purpose of information compression, thus $O(|\mathcal{V}|\times d^2) < O(|\mathcal{E}|\times d)$ empirically. In conclusion, the overall time complexity of \model\ is $O(|\mathcal{M}|\times |\mathcal{E}| \times d^2)$.\\\vspace{-0.12in}


\subsubsection{Space Complexity Analysis}
Although our \model\ uses latent memory blocks to enhance heterogeneous relation modeling, the extra parameters only costs $O(|\mathcal{M}|\times d^2)$ space for storing the learnable parameters. \model\ also requires additional $O(|\mathcal{E}|\times d^2)$ memory space for the edge-specific transformation matrices (\ie~$\sum_{m=1}^{|\mathcal{M}|}\eta(\textbf{H}^{(l)}[t], m)\textbf{W}_m^1$). In comparison, a standard GNN model requires $O(|\mathcal{V}|\times d)$ space to store node features. Also, $O(|\mathcal{E}|\times d)$ or $O(|\mathcal{E}|\times d^2)$ extra space is needed due to the edge-specific parameter customization 
(\eg~GAT~\cite{velivckovic2018graph}, HGT~\cite{wang2019heterogeneous}).



\begin{algorithm}[t]
    \caption{The Proposed \model\ Algorithm}
    \label{alg:model}
    \LinesNumbered
    \KwIn{user set $U=\{u_i\}$, item set $V=\{v_j\}$, relation set $R=\{r\}$, user-item interaction matrix $\textbf{Y}\in\mathbb{R}^{I\times J}$, user-user social matrix $\textbf{S}\in\mathbb{R}^{I\times I}$, item-item relational matrix $\textbf{T}\in\mathbb{R}^{J\times |R|}$, learning rate $\eta$, and number of epochs $E$}
    \KwOut{trained model parameters $\mathbf{\Theta}$}
    Initialize all parameters in $\mathbf{\Theta}$;\\
    Construct the collaborative heterogeneous graph $\mathcal{G}=\{\mathcal{V}, \mathcal{E}\}$ based on the relation matrices $\textbf{Y}, \textbf{S}, \textbf{T}$;\\
    \For{$e=1$ to $E$} {
        \For{$l=1$ to $L$} {
            \For{$(s, t)\in\mathcal{E}$} {
                Calculate the propagated information $\varphi(\textbf{H}^{(l)}[t], \textbf{H}^{(l)}[s])$ based on the memory-augmented encoder (Eq~\ref{eq:memoatt});
            }
            \For{$u_i\in U$} {
                Acquire aggregated information $\textbf{H}^{(l+1)}[u_i]$ from social and item relations (Eq~\ref{eq:userAgg});
            }
            \For{$v_j\in V$} {
                Calculate item embedding $\textbf{H}^{(l+1)}[v_j]$ based on user and item-relation edges (Eq~\ref{eq:itemAgg});
            }
            \For{$r\in R$} {
                Update the relation embedding $\textbf{H}^{(l+1)}[r]$ from the connected items (Eq~\ref{eq:relAgg});
            }
            Incorporate layer normalization and self-propagation to get $\widetilde{\textbf{H}}^{(l+1)}$ (Eq~\ref{eq:layerNorm});\\
        }
        Perform cross-layer embedding aggregation for $\textbf{H}^*$;\\
        Calculate loss $\mathcal{L}$ for a training batch (Eq~\ref{eq:loss});\\
        \For{$\theta\in\mathbf{\Theta}$} {
            $\theta =\theta - \eta\cdot \partial\mathcal{L}/\partial\theta$;
        }
    }
    \Return traind model with parameters $\mathbf{\Theta}$.
\end{algorithm}

\section{Evaluation}
\label{sec:eval}

In this section, we perform extensive experiments on three public real-world datasets for model performance evaluation by answering the research questions presented below:


\begin{itemize}[leftmargin=*]

\item \textbf{RQ1}: Compared with various state-of-the-art models, how does \model\ perform for making recommendations? \\\vspace{-0.12in}

\item \textbf{RQ2}: What is the impact of major components in \model? \\\vspace{-0.12in}

\item \textbf{RQ3}: How does different relation types (collaborative relations, social ties, and item-wise relations) contribute to the model performance of \model? \\\vspace{-0.12in}

\item \textbf{RQ4}: How does \model\ perform compared with baselines for user preference learning under data scarcity? \\\vspace{-0.12in}

\item \textbf{RQ5}: How do the key hyperparameters of \model\ model impact its performance with different settings? \\\vspace{-0.12in}

\item \textbf{RQ6}: How is the efficiency of our \model\ in both optimization phase and forecasting phase? \\\vspace{-0.12in}

\item \textbf{RQ7}: With the embedding visualization, how do the learned latent representations benefit from the collectively encoding of relation heterogeneity, from the social- and knowledge-enhanced user-item interactive patterns?

\end{itemize}

\begin{table}[t!]
\centering
\caption{Statistics of Experimented Datasets.}
\vspace{-0.05 in}
\begin{tabular}{l| c| c| c}
\hline
Dataset & Ciao & Epinions & Yelp\\
\hline
\# of Users & 1,925 & 18,081 & 99,262 \\
\# of Items & 15,053 & 251,722 & 10,5142 \\
\# of User-Item Interactions & 30,370 & 715,821 & 769,929 \\
Interaction Density Degree & 0.1048\% & 0.0157\% & 0.0074\% \\
\# of Social Ties & 65,084 & 572,784 & 1,298,522 \\
Social Tie Density Degree & 1.7564\% & 0.1752\% & 0.0132\% \\
\hline
\end{tabular}
\vspace{-0.1in}
\label{tab:data}
\end{table}

\subsection{Experimental Settings}

\subsubsection{\bf Datasets} Our experiments are conducted on three real-world benchmark datasets for social recommendation: \emph{Ciao}, \emph{Epinions} and \emph{Yelp}. These datasets are collected from different online review systems in real-life applications, where users can write reviews on different products. Furthermore, users can establish their social relationships by adding others into their trust lists. When constructing user-user connection network, an edge $e_{i,j}$ is added when user $u_i$ trust $u_j$ and vice versa. We summarize the statistics of our evaluation datasets in Table~\ref{tab:data}. Following the settings in~\cite{wang2019kgat,xin2019relational}, we generate the item-wise relations with external knowledge (\eg, product categories, business genres) from the item side.

\begin{table*}[t!]
\scriptsize
\centering
\caption{Performance comparison of all methods in terms of \emph{HR@10} and \emph{NDCG@10}. ``Imp'' represents the relatively performance improvement between our DHGM network and each compared baseline.} 
\vspace{-0.05in}
\setlength{\tabcolsep}{0.9mm}
\begin{tabular}{ c || c| c | c | c | c | c | c | c | c | c | c | c | c | c | c || c}
\hline
Dataset & Metrics & SAMN & EATNN & DiffNet & GraphRec & NGCF& GCCF & DGRec & KGAT & DGCF & DisenHAN & ~HAN~ & HGT & HERec & MHCN & ~\model ~ \\
\hline
\multirow{4}{*}{Ciao} & HR    & 0.4677 & 0.4130 & 0.5202 & 0.4594 & 0.4843 & 0.4926 & 0.5086 & 0.4907 & 0.5189 & 0.4856 & 0.4856 & 0.4933 & 0.5298 & 0.5080 & \textbf{0.5515} \\
\cline{2-17}
                    & Imp & 17.92\%  & 33.54\%  & 6.02\%  & 20.05\%  & 13.88\%  & 11.96\%  & 8.43\%  & 12.39\%  & 6.28\% & 13.57\% & 13.57\%  & 11.80\%  & 4.10\%  & 8.56\% & --\\
\cline{2-17}
                      & NDCG  & 0.2838 & 0.2520 & 0.3201 & 0.2670 & 0.3088 & 0.3070 & 0.3113 & 0.2977 & 0.3166 & 0.2894 & 0.2608 & 0.3062 & 0.3104 & 0.3118 & \textbf{0.3338}\\
                      \cline{2-17}
                      & Imp & 17.62\%  & 32.46\%  & 4.28\%  & 25.02\%  & 8.10\%  & 8.73\%  & 7.23\%  & 12.13\%  & 5.43\% & 15.34\% & 27.99\%  & 9.01\%  & 7.54\%  & 7.06\% & --\\

\cline{1-17}
\multirow{4}{*}{Epinions} & HR   & 0.6390 & 0.6422 & 0.6323 & 0.6865 & 0.6944 & 0.6779 & 0.6268 & 0.6756 & 0.6635 & 0.6825 & 0.6673 & 0.7001 & 0.6767 & 0.6411 & \textbf{0.7335} \\
\cline{2-17}
                        & Imp & 14.79\%  & 14.22\%  & 16.01\%  & 6.85\%  & 5.63\%  & 8.20\%  & 17.02\%  & 8.57\%  & 10.55\% & 7.47\% & 9.92\%  & 4.77\%  & 8.39\%  & 14.41\% & --\\
\cline{2-17}
                          & NDCG & 0.4259 & 0.4483 & 0.4160 & 0.4786 & 0.4763 & 0.4783 & 0.4127 & 0.4708 & 0.4594 & 0.4627 & 0.4371 & 0.4812 & 0.4572 & 0.4261 & \textbf{0.5215}\\
                          \cline{2-17}
                          & Imp  & 22.45\%  & 16.33\%  & 25.36\%  & 8.96\%  & 9.49\%  & 9.03\%  & 26.36\%  & 10.77\%  & 13.52\% & 12.71\% &  19.31\%  & 8.37\%  & 14.06\%  & 22.39\% & --\\
\hline
\multirow{4}{*}{Yelp} & HR       & 0.7971 & 0.7273 & 0.8222 & 0.8019 & 0.8204 & 0.8130 & 0.7830 & 0.7737 & 0.7956 & 0.8159 & 0.8169 & 0.8185 & 0.7047 & 0.8019 & \textbf{0.8373} \\
\cline{2-17}
                    & Imp & 5.04\%  & 15.12\%  & 1.84\%  & 4.41\%  & 2.06\%  & 2.99\%  & 6.93\%  & 8.22\%  & 5.24\% & 2.62\% & 2.50\%  & 2.30\%  & 18.82\%  & 4.41\% & --\\
\cline{2-17} 
                      & NDCG    & 0.5293 & 0.5289 & 0.5524 & 0.5372 & 0.5651 & 0.5585 & 0.5386 & 0.5386 & 0.5410 & 0.5403 & 0.5511 & 0.5547 & 0.4990 & 0.5348 & \textbf{0.5873} \\
                      \cline{2-17}
                      & Imp & 10.96\%  & 11.04\%  & 6.32\%  & 9.33\%  & 3.93\%  & 5.16\%  & 9.04\%  & 9.04\%  & 8.56\%  & 8.70\% & 6.57\%  & 5.88\%  & 17.70\%  & 9.82\% & --\\
\cline{1-17}
\hline
\end{tabular}
\label{tab:result}
\end{table*}

\subsubsection{\bf Compared Baselines} 
For comprehensive evaluation of model effectiveness, we compare \model\ with state-of-the-art methods from various research lines, covering i) attentive social recommendation models (SAMN, EATNN), ii) GNN-based social recommender systems (GraphRec, DiffNet, MHCN), iii) social recommendation with temporal context (DGRec), iv) neural graph collaborative filtering (NGCF, GCCF), v) disentangled graph recommender systems (DGCF, DisenHAN), vi) knowledge-aware recommendation model (KGAT), vii) heterogeneous graph representation learning for recommendation (HAN, HERec, HGT). \\\vspace{-0.12in}



\noindent \textbf{Attentive Social Recommender Systems}: Attention mechanisms have been serving as effective techniques to identify important relations for social-aware recommendation.
\begin{itemize}[leftmargin=*]
\item \textbf{SAMN}~\cite{chen2019social}: it designs a dual-stage attentional model to characterize the user-wise influence and select relevant friends to model user preference. The friend-wise attention is introduced to investigate the social influence among users. \\\vspace{-0.12in}

\item \textbf{EATNN}~\cite{chen2019efficient}: it is a transfer learning approach which fuses interaction and social information using attention mechanisms. An optimization scheme is designed to enable the multi-task learning framework.
\end{itemize}

\noindent \textbf{GNN-based Social Recommendation Models}: Graph neural networks have been utilized to model the graph-based user-item relationships and users' social ties for recommendation.
\begin{itemize}[leftmargin=*]
\item \textbf{DiffNet}~\cite{wu2019neural}: it uses the graph information propagation paradigm to model social relations with a layer-wise diffusion architecture for users. The dynamic social diffusion simulates the recursive social influence. \\\vspace{-0.12in}

\item \textbf{GraphRec}~\cite{fan2019graph}: it is built upon the graph attention network to propagate embeddings over social network, to enhance the representation of users. The social connection and item interactions are aggregated for user latent representations through the attentive combinations. \\\vspace{-0.12in}

\item \textbf{MHCN}~\cite{yu2021self}: it is a self-supervised learning architecture to capture user relationships using multi-channel hypergraph neural networks. The mutual information between the node-level and sub-graph-level embeddings are maximized, which serve as the auxiliary self-supervised learning task for joint training together with the recommendation loss.

\end{itemize}

\noindent \textbf{Graph Collaborative Filtering Models}: Graph collaborative filtering techniques become the effective recommendation solution by showing their state-of-the-art performance to capture the collaborative effects over user-item interaction graph. For fair comparison, we enhance the graph CF baselines by incorporating the diverse context into the interaction graph.

\begin{itemize}[leftmargin=*]
\item \textbf{GCCF}~\cite{chen2020revisiting}: it is a simplified GNN-based CF model with the utilization of convolution-based message passing. The non-linear transformation is removed from the graph convolutional network to address the overfitting issue. \\\vspace{-0.12in}


\item \textbf{NGCF}~\cite{wang2019neural}: it is a state-of-the-art graph convolution-based collaborative filtering model. In the representation process of users, the high-order connectivity is considered to inject collaborative signals into the user preference learning paradigm with recursively applying graph propagation functions.

\end{itemize}

\noindent \textbf{Temporal-aware Social Recommendation}: Another line of social recommendation lies in the incorporation of temporal context into the modeling of user-wise social influence.
\begin{itemize}[leftmargin=*]

\item \textbf{DGRec}~\cite{song2019session}: it incorporates the temporal context into the social recommendation with the integration of recurrent units and graph neural networks. Social connections are incorporated into dynamic interest representations of users.
\end{itemize}

\noindent \textbf{Disentangled Graph Recommender Systems}: Our \model\ model competes with the representative solutions with disentangled learning techniques for recommendation.
\begin{itemize}[leftmargin=*]

\item \textbf{DGCF}~\cite{wang2020disentangled}: this approach first partitions user embeddings into disjoint parts representing disentangled intent of user preference. Then, it perform intent-aware message passing over graph convolutional network for recommendation. \\\vspace{-0.12in}

\item \textbf{DisenHAN}~\cite{wang2020disenhan}: This model is built over the graph attention model to encode the disentangled embeddings based on different connections among users and items for propagation.

\end{itemize}

\noindent \textbf{Knowledge-aware Recommender System}:We also compare our proposed \model\ framework with the recommendation algorithm which utilizes the knowledge graph as the item side information to learn the item semantic relatedness.
\begin{itemize}[leftmargin=*]
\item \textbf{KGAT}~\cite{wang2019kgat}: it is a knowledge-enhanced model using attention to aggregate information from both user-item interactions and item knowledge graph. The weights of neighboring nodes and entities are learned through the attention layer.
\end{itemize}

\noindent \textbf{Heterogeneous Graph Representation for Recommendation}: In the performance comparison, we include two state-of-the-art heterogeneous graph embedding techniques to model the different relationships in recommender systems.

\begin{itemize}[leftmargin=*]

\item \textbf{HAN}~\cite{wang2019heterogeneous}: it encodes the heterogeneity of graph with a meta-path-guided hierarchical attention model consisting of node- and semantic-level attention. We apply this method to encode the node representations in our collaborative heterogeneous graph $\mathcal{G}$, to preserve both the social-aware user dependencies and knowledge-aware item correlations. \\\vspace{-0.12in}

\item \textbf{HGT}~\cite{wang2019heterogeneous}: this method is a graph transformer architecture for heterogeneous graph representation learning. It calculates edge-specific transformation and attention for graph message passing with heterogeneity encoding. \\\vspace{-0.12in}

\item \textbf{HERec}~\cite{shi2018heterogeneous}: it is a heterogeneous network embedding method which integrates various fusion functions with meta-path random walk strategy, to incorporate various side information into the user preference learning.
\end{itemize}

\begin{table*}[t!]
\vspace{-0.1in}
\caption{Performance evaluation with varying Top-\textit{N} in terms of \textit{HR@N} and \textit{NDCG@N}.}
\vspace{-0.1in}
 \centering
\footnotesize
 \begin{tabular}{|c|c|c|c|c|c|c|c|c|c|c|c|c|}
  \hline
  \multirow{2}{*}{Model}&\multicolumn{2}{c|}{Ciao@5}&\multicolumn{2}{c|}{Ciao@20}&\multicolumn{2}{c|}{Epinions@5}&\multicolumn{2}{c|}{Epinions@20}&\multicolumn{2}{c|}{Yelp@5}&\multicolumn{2}{c|}{Yelp@20}\\
  \cline{2-13}
  & HR & NDCG & HR & NDCG & HR & NDCG & HR & NDCG & HR & NDCG & HR & NDCG\\
  \hline
  \hline
        SAMN & 0.3468 & 0.2460 & 0.6251 & 0.3223 & 0.5176 & 0.3860 & 0.7491 & 0.4553 & 0.6359 & 0.4662 & 0.9009 & 0.5407\\
        \hline
        EATNN & 0.2969 & 0.2124 & 0.5222 & 0.2819 & 0.5283 & 0.3924 & 0.7501 & 0.4557 & 0.6425 & 0.4866 & 0.8066 & 0.5468\\
        \hline
        DiffNet & 0.3941 & 0.2816 & 0.6647 & 0.3573 & 0.5106 & 0.3820 & 0.7367 & 0.4476 & 0.6701 & 0.5127 & 0.9053 & 0.5701 \\
        \hline
        GraphRec & 0.3058 & 0.2235 & 0.5976 & 0.3042 & 0.5683 & 0.4325 & 0.8001 & 0.5011 & 0.6631 & 0.4903 & 0.8944 & 0.5650\\
        \hline
        NGCF & 0.3570 & 0.2360 & 0.5937 & 0.3188 & 0.5612 & 0.4316 & 0.8010 & 0.5006 & 0.6748 & 0.5192 & 0.9011 & 0.5684 \\
        \hline
        GCCF & 0.3685 & 0.2668 & 0.6289 & 0.3414 & 0.5538 & 0.4161 & 0.7906 & 0.4852 & 0.6703 & 0.5130 & 0.9011 & 0.5503 \\
        \hline
        DGRec & 0.3724 & 0.2647 & 0.6219 & 0.3277 & 0.5053 & 0.3775 & 0.7308 & 0.4429 & 0.6511 & 0.4897 & 0.8824 & 0.5611 \\
        \hline
        KGAT & 0.3391 & 0.2422 & 0.6052 & 0.3326 & 0.5483 & 0.4139 & 0.7880 & 0.4837 & 0.6503 & 0.4901 & 0.8795 & 0.5521\\
        \hline
        DGCF & 0.3871 & 0.2782 & 0.6775 & 0.3604 & 0.5479 & 0.4144 & 0.7770 & 0.4811 & 0.6565 & 0.4958& 0.9010 & 0.5678\\
        \hline
        DisenHAN & 0.3493 & 0.2482 & 0.6161 & 0.3248 & 0.5609 & 0.4247 & 0.7890 & 0.4911 & 0.6511 & 0.4944 & 0.9040 & 0.5650\\
        \hline
        HAN & 0.2937 & 0.1897 & 0.6513 & 0.2821 & 0.5403 & 0.4106 & 0.7761 & 0.4802 & 0.6635 & 0.5080 & 0.8977 & 0.5529\\
        \hline
        HGT  & 0.3415 & 0.2372 & 0.6128 & 0.3229 & 0.5757 & 0.4360 & 0.8053 & 0.5029 & 0.6888 & 0.5136 & 0.9060 & 0.5802 \\
        \hline
        HERec & 0.3832 & 0.2679 & 0.6846 & 0.3641 & 0.5519 & 0.4179 & 0.7792 & 0.4839 & 0.5833 & 0.4501 & 0.8125 & 0.5034\\
        \hline
        MHCN  & 0.3864 & 0.2799 & 0.6321 & 0.3453 & 0.5199 & 0.3883 & 0.7496 & 0.4551 & 0.6607 & 0.4911 & 0.8958 & 0.5670 \\
        \hline
        \hline
        \model\ & \textbf{0.4120} & \textbf{0.2890} & \textbf{0.6942} & \textbf{0.3726} & \textbf{0.6142} & \textbf{0.4794} & \textbf{0.8281} & \textbf{0.5387} & \textbf{0.7052} & \textbf{0.5378} & \textbf{0.9293} & \textbf{0.6043} \\
  \hline
 \end{tabular}
 \label{tab:vary_k}
\end{table*}

\subsubsection{\bf Evaluation Metrics}
We focus on top-$N$ item recommendation and use two widely adopted metrics for evaluation: Hit Rate (HR)@$N$ and Normalized Discounted Cumulative Gain (NDCG)@$N$ with the top-$N$ ranked positions~\cite{chen2018sequential, wang2019neural}, to measure the recommendation accuracy of each evaluated method. For individual target user, we select 100 non-interacted items as negative samples and combine them with the interacted item (as positive instances) in the evaluation procedure. Formally, the metrics are calculated as follows:
\begin{align}
HR@N&=\frac{\sum_{i=1}^M\sum_{j=1}^N r_{i,j}}{M}\nonumber\\
NDCG@N&=\sum_{i=1}^M \frac{\sum_{j=1}^N r_{i,j} / \log_2(j+1)}{M\cdot IDCG_i}
\end{align}
\noindent where $M$ denotes the number of tested users. $r_{i,j}=1$ if the $j$-th item in the ranked list of the $i$-th user is the positive item, and $r_{i,j}=0$ otherwise. The numerator of NDCG@$N$ is the discounted cumulative gain (DCG)@$N$, and $IDCG_i$ denotes the possible maximum DCG@$N$ value for the $i$-th tested user.
\\\vspace{-0.1in}

\subsubsection{\bf Hyperparameter Settings}
We implement the proposed \model\ framework based on Pytorch and perform the model optimization with Adam. For our \model\ method, the dimensionality of embedding is tuned from the range [4, 8, 16, 32]. The learning rate is set as $0.01$ and the batch size is searched between 512 and 4096. The coefficient $\lambda$ of regularization term is tuned in \{$10^{-3}, 10^{-4}, 10^{-5}$\}. In our experiments, we set the number of latent memory units as 8. Other hyperparameter details can be found in our release source code.



\subsection{Performance Comparison (RQ1)}
The empirical results of all compared methods on three different datasets (\ie, Ciao, Epinions, Yelp) are reported in Table~\ref{tab:result}. We summarize the following major observations:\\\vspace{-0.12in}

Our \model\ achieves the best performance as compared to all baselines across three different datasets, which demonstrates the performance superiority of \model. Such performance improvements are attributed to the following model design: i) Benefiting from our heterogeneous graph memory network, \model\ could preserve the comprehensive relation semantics with latent factor disentanglement, which results in the effectively integration of disentangled social- and knowledge-aware collaborative signals. i) By incorporating knowledge-aware item relations into social recommendation framework, \model\ can better characterize the heterogeneous relationships among users and items, which enhances the representation paradigm of user-item interactions.\\\vspace{-0.1in}

GNN-based social recommendation models perform better than the attentional solutions, which suggests the rationality of performing the embedding propagation with multi-hop graph structures for social relation transformation. In addition, although the design of transforming various relationships via heterogeneous graph encoders in HAN and HERec, they are limited by the generation of meta-path-guided relations with data-specific domain knowledge. To address this issue, \model\ learns more powerful node- and edge-type dependent representations by avoiding customized meta paths in the embedding function of relation heterogeneity.\\\vspace{-0.1in}

The performance gap between \model\ and GNN-based methods (\eg, DiffNet, GraphRec, DGRec), indicates that leaving the knowledge-aware item relations untapped will limit the performance of social-aware recommender systems. Such observation also suggests that \model\ is good at fulfilling the potentials of learning latent factors from both user and item domains, with the designed memory-enhanced graph neural architecture. From Table~\ref{tab:vary_k}, we can observe the performance gain achieved by \model\ over other competitors with different ranked top-$N$ positions, which further justifies the superior ranking performance of our framework. The recommendation accuracy improves with larger $N$ values.\\\vspace{-0.1in}

\subsection{Module Ablation Analyses (RQ2)}

\begin{figure}
    \centering
    \subfigure[][Ciao-HR]{
        \centering
        \includegraphics[width=0.27\columnwidth]{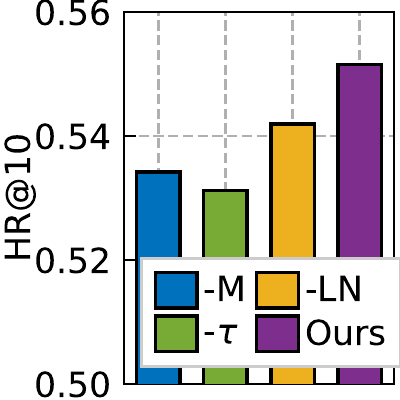}
    }
    \subfigure[][Epinions-HR]{
        \centering
        \includegraphics[width=0.27\columnwidth]{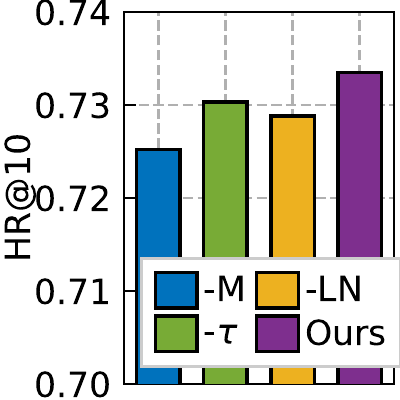}
    }
    \subfigure[][Yelp-HR]{
        \centering
        \includegraphics[width=0.27\columnwidth]{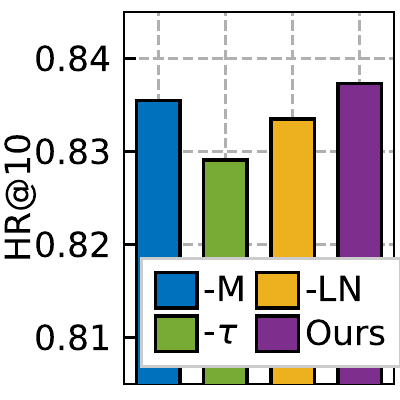}
    }
    \subfigure[][Ciao-NDCG]{
        \centering
        \includegraphics[width=0.27\columnwidth]{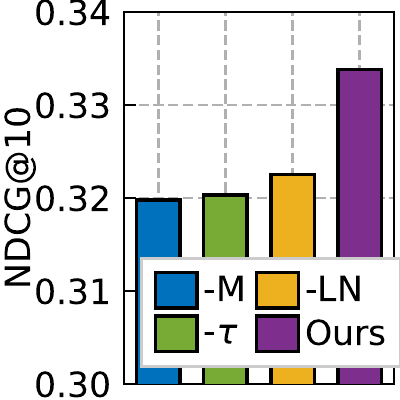}
    }
    \subfigure[][Epinions-NDCG]{
        \centering
        \includegraphics[width=0.27\columnwidth]{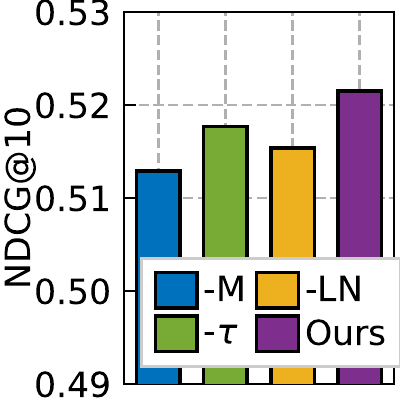}
    }
    \subfigure[][Yelp-NDCG]{
        \centering
        \includegraphics[width=0.27\columnwidth]{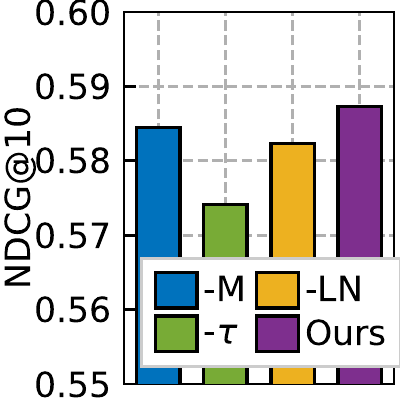}
    }
    \caption{Ablation studies for different sub-modules in \model\ framework on Ciao, Epinions and Yelp datasets, in terms of HR@10 and NDCG@10.}
    \vspace{-0.2in}
    \label{fig:mod_ablation}
\end{figure}

In this section, we investigate the design rationality of sub-networks in our \model\ framework. Towards this end, we remove each of key modules and implement three model variants of \model\ corresponding to three technical points of our \model: i) ``-M'': \model\ without the disentangled memory-enhanced relation heterogeneity encoder. ii) ``-$\tau$'': \model\ without the incorporation of user-specific social influence with the representation recalibration function $\tau(\cdot)$. iii) ``-LN'': \model\ without the layer normalization in each propagation layer for stable node embedding training.

We evaluate the performance of the above variants as well as \model\ on three experimental data. The results are shown in Figure~\ref{fig:mod_ablation}. \model\ consistently achieves best performance in comparison to the three variants. By inspecting the results in detail, we have the following observations: i) Removing the memory-enhanced relation heterogeneity encoder (\ie, ``-M'') causes significant performance degradation. This validates the effectiveness of disentangling latent factors pertinent to each type of relations. ii) The performance gap between the ``-$\tau$'' variant and \model\ indicates that using local structures in social relations benefits the user representation learning through explicitly involving heterogeneous relation data. iii) By comparing ``-LN'' with \model, we can conclude that the layer normalization technique has contribution to the model training of \model. We ascribe the improvements to its ability to generate stable gradients through normalization.

\subsection{Effect of Heterogeneous Relationships (RQ3)}
We further investigate the influence of different auxiliary relational data (\ie, social connections and item-wise relatedness) on the performance of our \model. In specific, the following three model variants are considered: i) ``-T'': \model\ removes the item relation matrix $\textbf{T}\in\mathbb{R}^{J\times |R|}$. ii) ``-S'': \model\ without the user-user social relation matrix $\textbf{S}\in\mathbb{R}^{I\times I}$ in the joint adjacent matrix. iii) ``-ST'': both user-wise social relations and item-wise relations are removed from the input.

\begin{figure}
    \centering
    \subfigure[][Ciao-HR] {
        \centering
        \includegraphics[width=0.46\columnwidth]{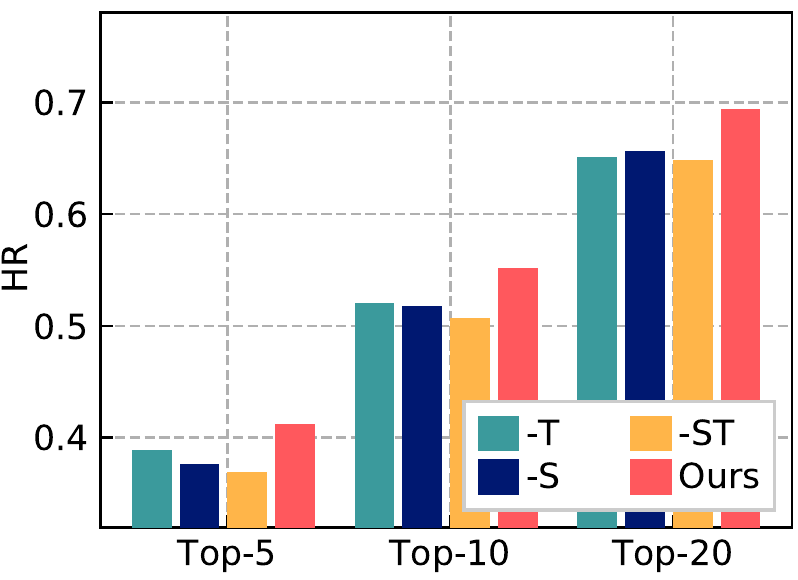}
    }
    \subfigure[][Ciao-NDCG] {
        \centering
        \includegraphics[width=0.46\columnwidth]{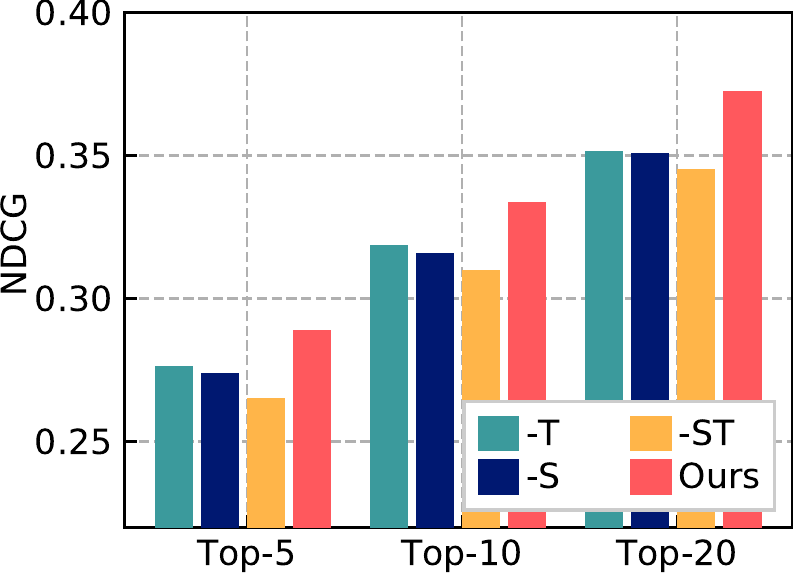}
    }
    \subfigure[][Epinions-HR] {
        \centering
        \includegraphics[width=0.46\columnwidth]{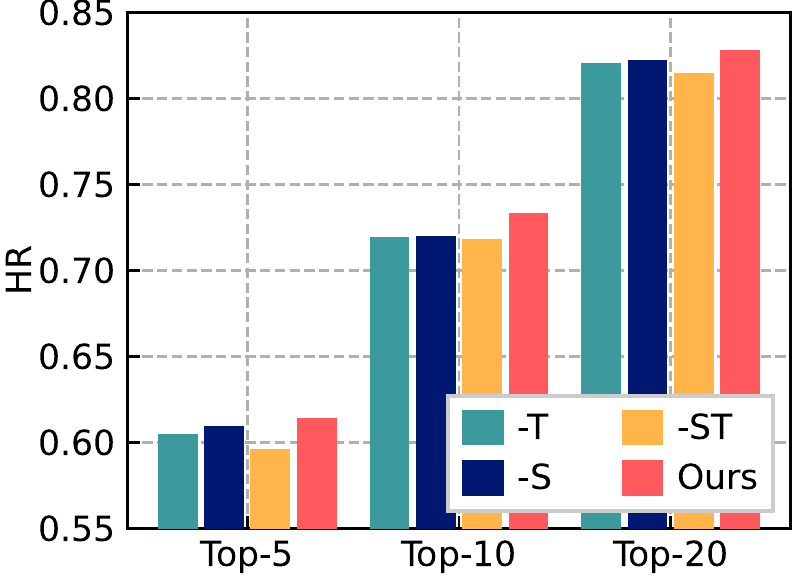}
    }
    \subfigure[][Epinions-NDCG] {
        \centering
        \includegraphics[width=0.46\columnwidth]{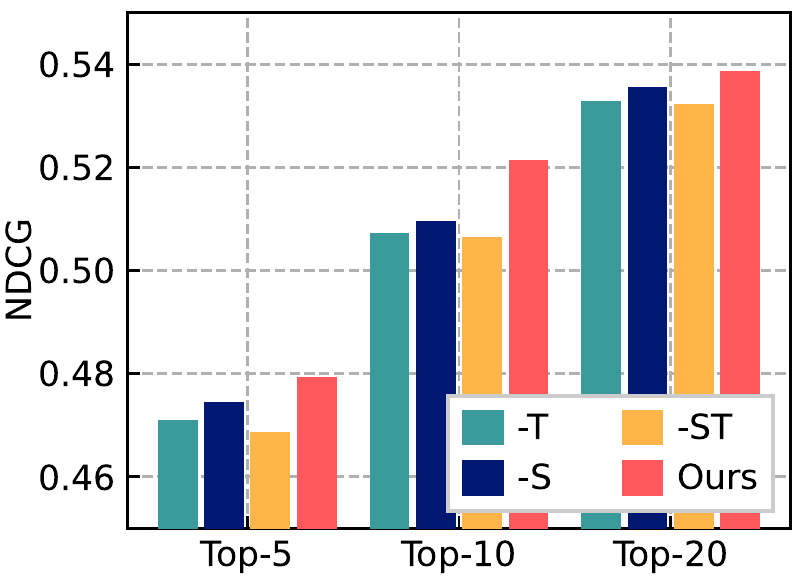}
    }
    \subfigure[][Yelp-HR] {
        \centering
        \includegraphics[width=0.46\columnwidth]{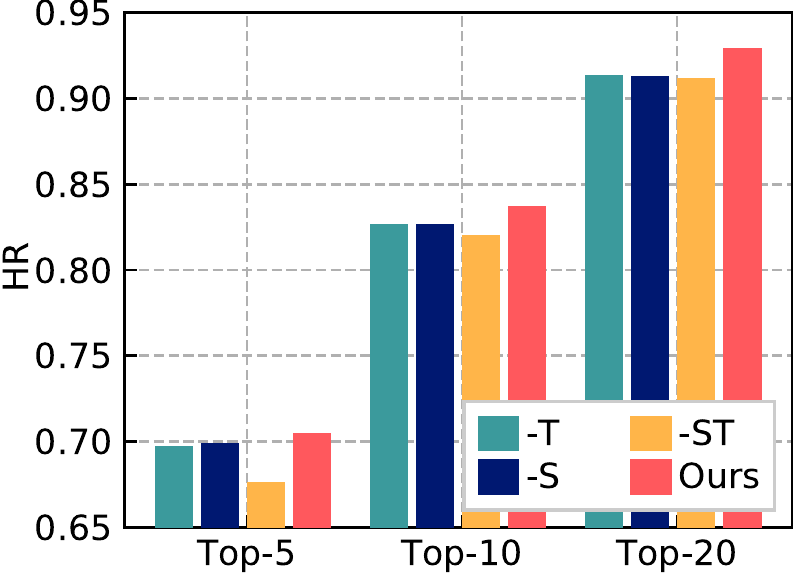}
    }
    \subfigure[][Yelp-NDCG] {
        \centering
        \includegraphics[width=0.46\columnwidth]{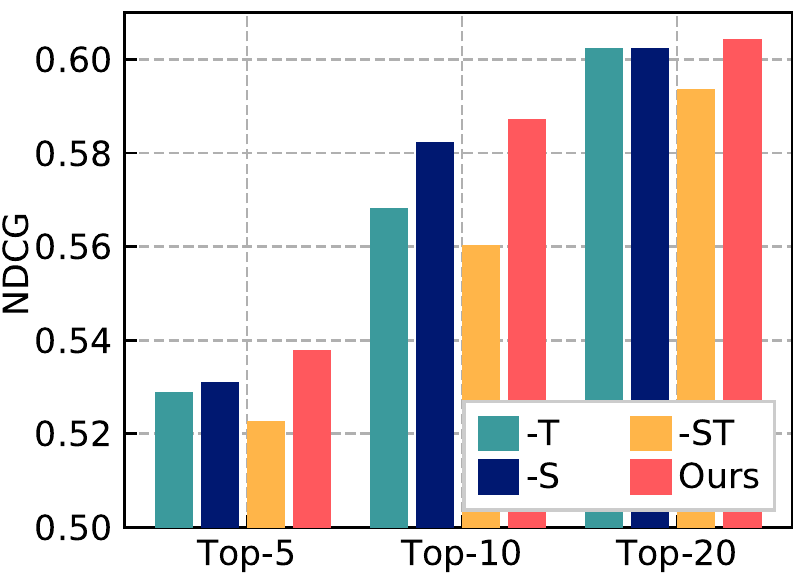}
    }
    \caption{Ablation studies on the effect of different heterogeneous data on model performance, in terms of HR@N and NDCG@N (N=5, 10, 20).}
    \vspace{-0.1in}
    \label{fig:data_ablation}
\end{figure}

The evaluation is conducted on Ciao data and Yelp data, with varying top-N settings. The results are shown in Figure~\ref{fig:data_ablation}. Major conclusions can be drawn as follows: 1) The auxiliary heterogeneous relations consistently bring positive effects to the model performance, which can be attributed to the benefit of incorporated heterogeneous semantics into the representations. 2) The suboptimal performance of ``-S'' suggests the helpfulness of leveraging social contextual signals to assist user preference learning. 3) \model\ performs better than ``-T'' in all cases. We ascribe the improvements to excavating the rich item-wise dependencies with our memory-enhanced relation heterogeneity encoder. 4) The ``-ST'' variant always produces the worst performance in both datasets, which further indicates that incorporating the heterogeneous relations from either user or item domains can improve the accuracy.


\subsection{Performance on Alleviating Data Sparsity (RQ4)}
\begin{figure}[t]
    \centering
    \subfigure[][\scriptsize{Interaction Factor-NDCG}]{
        \centering
        \includegraphics[width=0.45\columnwidth]{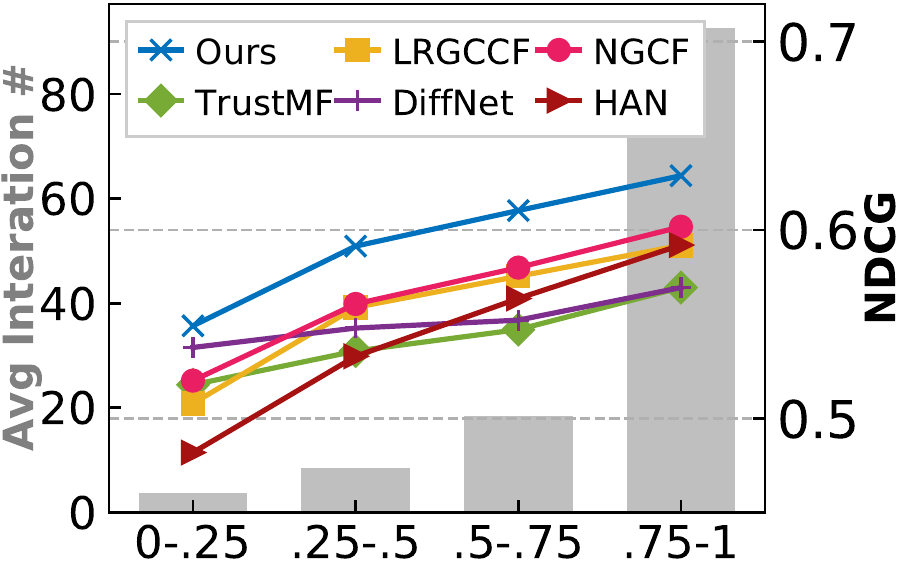}
        \label{sparsity_yelp_ndcg}
    }
    \subfigure[][\scriptsize{Interaction Factor-HR}]{
        \centering
        \includegraphics[width=0.45\columnwidth]{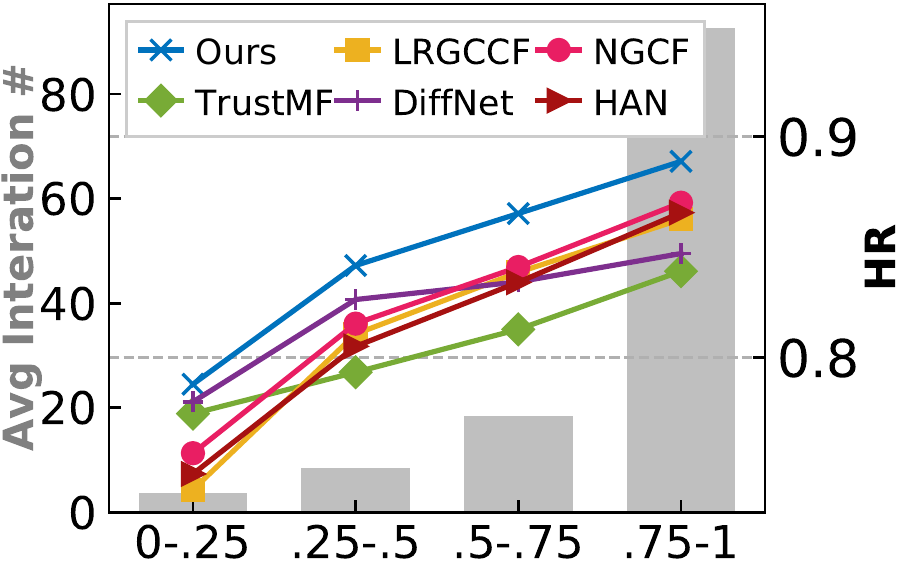}
        \label{sparsity_yelp_hr}
    }
    \subfigure[][\scriptsize{Social and Interaction Factors}]{
        \centering
        \includegraphics[width=0.45\columnwidth]{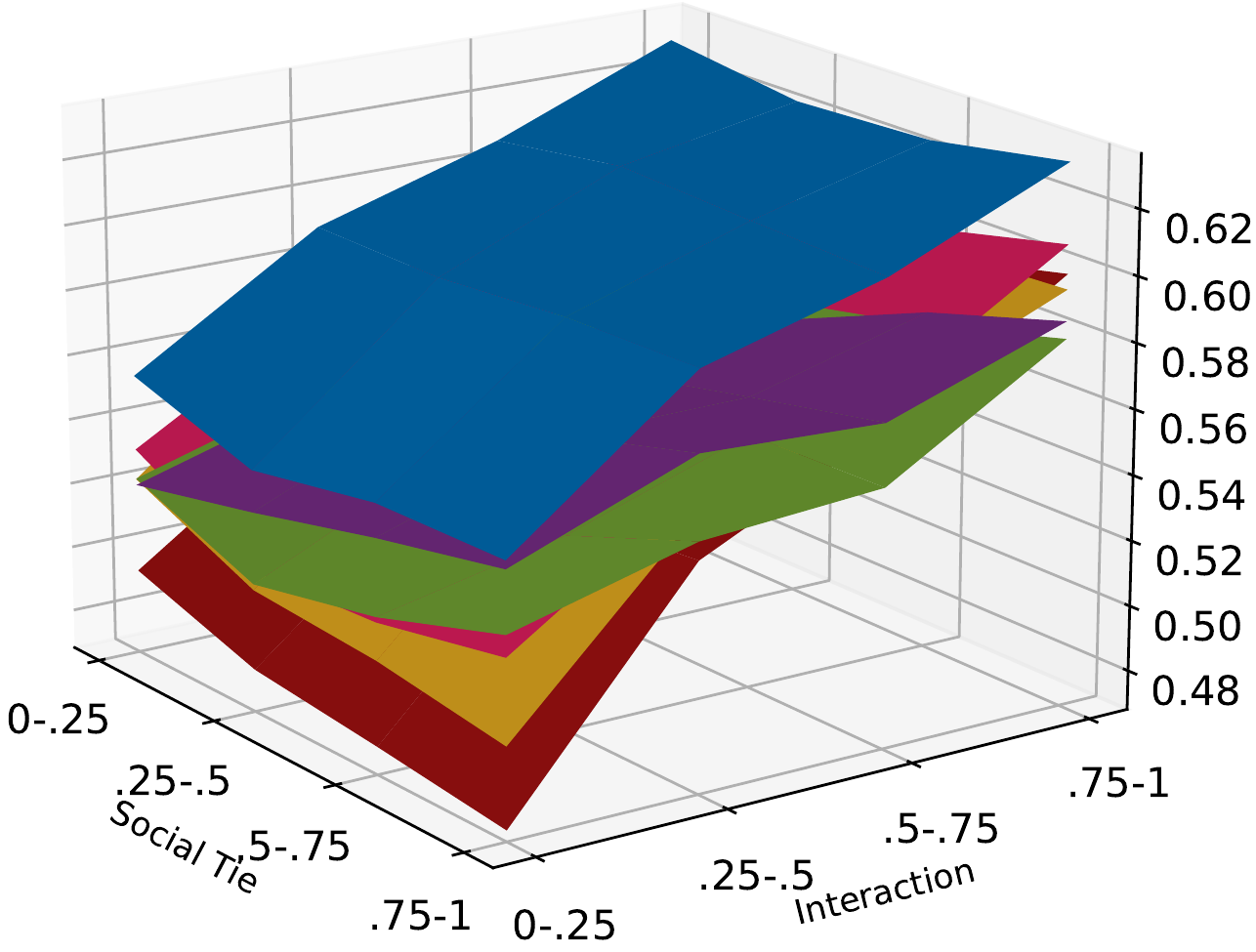}
        \label{hgmn_visualization}
    }
    \subfigure[][\scriptsize{Social Factor}]{
        \centering
        \includegraphics[width=0.45\columnwidth]{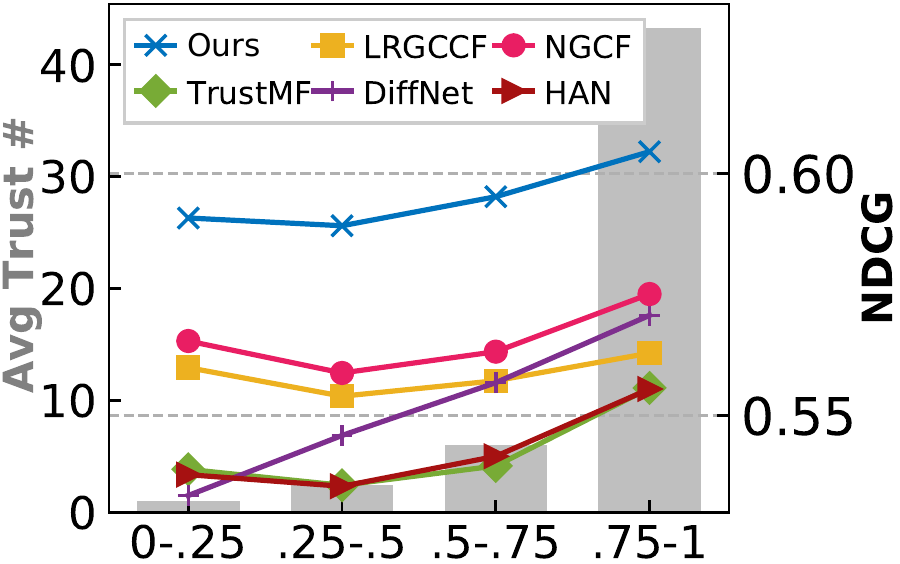}
        \label{ngcf_visualization}
    }
    \vspace{-0.05in}
    \caption{Performance comparison of \model\ and baselines
with different sparsity levels in terms of user interactions and social connections on Yelp.}
    \label{fig:sparsity}
\end{figure}

In this subsection, we perform experiments to demonstrate the advantage of our \model\ in incorporating heterogeneous side information from both user and item domain, in order to alleviate the sparsity issue of recommendation. We first rank all users in
terms of their interaction densities and partition them into four different groups which contains equal number of users (as shown in x-axis of Figure~\ref{fig:sparsity}: 0-25\%; 25\%-50\%, etc). In Figure~\ref{fig:sparsity}, we calculate the average number of interactions for each user group as shown in left side of y-axis. The recommendation performance of each compared method is presented in the right side of y-axis. From the results, we can observe that \model\ performs best compared with baselines on different datasets, which shows the robustness of our recommender system in dealing with the sparsity of user behavior data. This again further confirms that the effectiveness of \model\ for enabling external knowledge from both user and item domain to guide the user preference embedding with cross-relational context under data scarcity.

We further show the evaluation results \wrt\ two different factors (social and interaction dimensions). As shown in Figure~\ref{fig:sparsity}, we can observe that \model\ consistently outperforms competitive methods with different data sparsity levels, which justifies the effectiveness of \model\ in alleviating the sparsity from the perspectives of both social relations and user-item interactions. Overall, this property of our \model\ method in alleviating the data sparsity problem is important, for the recommendation scenarios, in which there are few user-item interactions compared with the entire interaction space.

\subsection{Hyperparameter Study (RQ5)}
\begin{figure*}
    \centering
    \begin{adjustbox}{max width=1.0\linewidth}
    \begin{tikzpicture}
\begin{axis}[
    xlabel={Hidden State Size $d$},
    ylabel={HR@10},
    xmin=2,xmax=34,
    ymin=0.4, ymax=0.56,
    legend columns=1,
    legend cell align=right,
    grid=both,
    every axis plot/.append style={ultra thick},
    every tick label/.append style={scale=1.2},
    label style={scale=2},
    legend style={
        nodes={scale=1.5, transform shape},
        legend image post style={scale=1.5},
        },
    legend style={at={(1,0)},anchor=south east},
    every axis plot post/.append style={
        every mark/.append style={scale=2.2}
    }
]
\addplot[color={rgb:red,4}, mark=o, dashed, mark options={solid}]
table[x=para, y=Ciao] {latdimHR.txt};
\legend{Ciao};
\end{axis}
\end{tikzpicture}

\begin{tikzpicture}
\begin{axis}[
    xlabel={Hidden State Size $d$},
    ylabel={NDCG@10},
    xmin=2,xmax=34,
    ymin=0.23, ymax=0.34,
    legend columns=1,
    legend cell align=right,
    grid=both,
    every axis plot/.append style={ultra thick},
    every tick label/.append style={scale=1.2},
    label style={scale=2},
    legend style={
        nodes={scale=1.5, transform shape},
        legend image post style={scale=1.5},
        },
    legend style={at={(1,0)},anchor=south east},
    every axis plot post/.append style={
        every mark/.append style={scale=2.2}
    }
]
\addplot[color={rgb:red,4}, mark=o, dashed, mark options={solid}]
table[x=para, y=Ciao] {latdimNDCG.txt};
\legend{Ciao};
\end{axis}
\end{tikzpicture}

\begin{tikzpicture}
\begin{axis}[
    xlabel={\# of GNN Layers $L$},
    ylabel={HR@10},
    xmin=-0.1,xmax=3.1,
    ymin=0.32, ymax=0.56,
    legend columns=1,
    legend cell align=right,
    grid=both,
    every axis plot/.append style={ultra thick},
    every tick label/.append style={scale=1.2},
    label style={scale=2},
    legend style={
        nodes={scale=1.5, transform shape},
        legend image post style={scale=1.5},
        },
    legend style={at={(1,0)},anchor=south east},
    every axis plot post/.append style={
        every mark/.append style={scale=2.2}
    }
]
\addplot[color={rgb:red,4}, mark=o, dashed, mark options={solid}]
table[x=para, y=Ciao] {gnnLayersHR.txt};
\legend{Ciao};
\end{axis}
\end{tikzpicture}

\begin{tikzpicture}
\begin{axis}[
    xlabel={\# of GNN Layers $L$},
    ylabel={NDCG@10},
    xmin=-0.1,xmax=3.1,
    ymin=0.15, ymax=0.35,
    legend columns=1,
    legend cell align=right,
    grid=both,
    every axis plot/.append style={ultra thick},
    every tick label/.append style={scale=1.2},
    label style={scale=2},
    legend style={
        nodes={scale=1.5, transform shape},
        legend image post style={scale=1.5},
        },
    legend style={at={(1,0)},anchor=south east},
    every axis plot post/.append style={
        every mark/.append style={scale=2.2}
    }
]
\addplot[color={rgb:red,4}, mark=o, dashed, mark options={solid}]
table[x=para, y=Ciao] {gnnLayersNDCG.txt};
\legend{Ciao};
\end{axis}
\end{tikzpicture}

\begin{tikzpicture}
\begin{axis}[
    xlabel={\# of Memory Units $|\mathcal{M}|$},
    ylabel={HR@10},
    xmin=1,xmax=17,
    ymin=0.48, ymax=0.56,
    legend columns=1,
    legend cell align=right,
    grid=both,
    every axis plot/.append style={ultra thick},
    every tick label/.append style={scale=1.2},
    label style={scale=2},
    legend style={
        nodes={scale=1.5, transform shape},
        legend image post style={scale=1.5},
        },
    legend style={at={(1,0)},anchor=south east},
    every axis plot post/.append style={
        every mark/.append style={scale=2.2}
    }
]
\addplot[color={rgb:red,4}, mark=o, dashed, mark options={solid}]
table[x=para, y=Ciao] {memoryUnitHR.txt};
\legend{Ciao};
\end{axis}
\end{tikzpicture}

\begin{tikzpicture}
\begin{axis}[
    xlabel={\# of Memory Units $|\mathcal{M}|$},
    ylabel={NDCG@10},
    xmin=1,xmax=17,
    ymin=0.27, ymax=0.34,
    legend columns=1,
    legend cell align=right,
    grid=both,
    every axis plot/.append style={ultra thick},
    every tick label/.append style={scale=1.2},
    label style={scale=2},
    legend style={
        nodes={scale=1.5, transform shape},
        legend image post style={scale=1.5},
        },
    legend style={at={(1,0)},anchor=south east},
    every axis plot post/.append style={
        every mark/.append style={scale=2.2}
    }
]
\addplot[color={rgb:red,4}, mark=o, dashed, mark options={solid}]
table[x=para, y=Ciao] {memoryUnitNDCG.txt};
\legend{Ciao};
\end{axis}
\end{tikzpicture}
    \end{adjustbox}
    \begin{adjustbox}{max width=1.0\linewidth}
    \begin{tikzpicture}
\begin{axis}[
    xlabel={Hidden State Size $d$},
    ylabel={HR@10},
    xmin=2,xmax=34,
    ymin=0.65, ymax=0.74,
    legend columns=1,
    legend cell align=right,
    grid=both,
    every axis plot/.append style={ultra thick},
    every tick label/.append style={scale=1.2},
    label style={scale=2},
    legend style={
        nodes={scale=1.5, transform shape},
        legend image post style={scale=1.5},
        },
    legend style={at={(1,0)},anchor=south east},
    every axis plot post/.append style={
        every mark/.append style={scale=2.2}
    }
]
\addplot[color={rgb:red,4}, mark=o, dashed, mark options={solid}]
table[x=para, y=Epinions] {latdimHR.txt};
\legend{Epinions};
\end{axis}
\end{tikzpicture}

\begin{tikzpicture}
\begin{axis}[
    xlabel={Hidden State Size $d$},
    ylabel={NDCG@10},
    xmin=2,xmax=34,
    ymin=0.44, ymax=0.53,
    legend columns=1,
    legend cell align=right,
    grid=both,
    every axis plot/.append style={ultra thick},
    every tick label/.append style={scale=1.2},
    label style={scale=2},
    legend style={
        nodes={scale=1.5, transform shape},
        legend image post style={scale=1.5},
        },
    legend style={at={(1,0)},anchor=south east},
    every axis plot post/.append style={
        every mark/.append style={scale=2.2}
    }
]
\addplot[color={rgb:red,4}, mark=o, dashed, mark options={solid}]
table[x=para, y=Epinions] {latdimNDCG.txt};
\legend{Epinions};
\end{axis}
\end{tikzpicture}

\begin{tikzpicture}
\begin{axis}[
    xlabel={\# of GNN Layers $L$},
    ylabel={HR@10},
    xmin=-0.1,xmax=3.1,
    ymin=0.57, ymax=0.745,
    legend columns=1,
    legend cell align=right,
    grid=both,
    every axis plot/.append style={ultra thick},
    every tick label/.append style={scale=1.2},
    label style={scale=2},
    legend style={
        nodes={scale=1.5, transform shape},
        legend image post style={scale=1.5},
        },
    legend style={at={(1,0)},anchor=south east},
    every axis plot post/.append style={
        every mark/.append style={scale=2.2}
    }
]
\addplot[color={rgb:red,4}, mark=o, dashed, mark options={solid}]
table[x=para, y=Epinions] {gnnLayersHR.txt};
\legend{Epinions};
\end{axis}
\end{tikzpicture}

\begin{tikzpicture}
\begin{axis}[
    xlabel={\# of GNN Layers $L$},
    ylabel={NDCG@10},
    xmin=-0.1,xmax=3.1,
    ymin=0.35, ymax=0.53,
    legend columns=1,
    legend cell align=right,
    grid=both,
    every axis plot/.append style={ultra thick},
    every tick label/.append style={scale=1.2},
    label style={scale=2},
    legend style={
        nodes={scale=1.5, transform shape},
        legend image post style={scale=1.5},
        },
    legend style={at={(1,0)},anchor=south east},
    every axis plot post/.append style={
        every mark/.append style={scale=2.2}
    }
]
\addplot[color={rgb:red,4}, mark=o, dashed, mark options={solid}]
table[x=para, y=Epinions] {gnnLayersNDCG.txt};
\legend{Epinions};
\end{axis}
\end{tikzpicture}

\begin{tikzpicture}
\begin{axis}[
    xlabel={\# of Memory Units $|\mathcal{M}|$},
    ylabel={HR@10},
    xmin=1,xmax=17,
    ymin=0.72, ymax=0.735,
    legend columns=1,
    legend cell align=right,
    grid=both,
    every axis plot/.append style={ultra thick},
    every tick label/.append style={scale=1.2},
    label style={scale=2},
    legend style={
        nodes={scale=1.5, transform shape},
        legend image post style={scale=1.5},
        },
    legend style={at={(1,0)},anchor=south east},
    every axis plot post/.append style={
        every mark/.append style={scale=2.2}
    }
]
\addplot[color={rgb:red,4}, mark=o, dashed, mark options={solid}]
table[x=para, y=Epinions] {memoryUnitHR.txt};
\legend{Epinions};
\end{axis}
\end{tikzpicture}

\begin{tikzpicture}
\begin{axis}[
    xlabel={\# of Memory Units $|\mathcal{M}|$},
    ylabel={NDCG@10},
    xmin=1,xmax=17,
    ymin=0.47, ymax=0.525,
    legend columns=1,
    legend cell align=right,
    grid=both,
    every axis plot/.append style={ultra thick},
    every tick label/.append style={scale=1.2},
    label style={scale=2},
    legend style={
        nodes={scale=1.5, transform shape},
        legend image post style={scale=1.5},
        },
    legend style={at={(1,0)},anchor=south east},
    every axis plot post/.append style={
        every mark/.append style={scale=2.2}
    }
]
\addplot[color={rgb:red,4}, mark=o, dashed, mark options={solid}]
table[x=para, y=Epinions] {memoryUnitNDCG.txt};
\legend{Epinions};
\end{axis}
\end{tikzpicture}
    \end{adjustbox}
    \begin{adjustbox}{max width=1.0\linewidth}
    \begin{tikzpicture}
\begin{axis}[
    xlabel={Hidden State Size $d$},
    ylabel={HR@10},
    xmin=2,xmax=34,
    ymin=0.76, ymax=0.84,
    legend columns=1,
    legend cell align=right,
    grid=both,
    every axis plot/.append style={ultra thick},
    every tick label/.append style={scale=1.2},
    label style={scale=2},
    legend style={
        nodes={scale=1.5, transform shape},
        legend image post style={scale=1.5},
        },
    legend style={at={(1,0)},anchor=south east},
    every axis plot post/.append style={
        every mark/.append style={scale=2.2}
    }
]
\addplot[color={rgb:red,4}, mark=o, dashed, mark options={solid}]
table[x=para, y=Yelp] {latdimHR.txt};
\legend{Yelp};
\end{axis}
\end{tikzpicture}

\begin{tikzpicture}
\begin{axis}[
    xlabel={Hidden State Size $d$},
    ylabel={NDCG@10},
    xmin=2,xmax=34,
    ymin=0.52, ymax=0.59,
    legend columns=1,
    legend cell align=right,
    grid=both,
    every axis plot/.append style={ultra thick},
    every tick label/.append style={scale=1.2},
    label style={scale=2},
    legend style={
        nodes={scale=1.5, transform shape},
        legend image post style={scale=1.5},
        },
    legend style={at={(1,0)},anchor=south east},
    every axis plot post/.append style={
        every mark/.append style={scale=2.2}
    }
]
\addplot[color={rgb:red,4}, mark=o, dashed, mark options={solid}]
table[x=para, y=Yelp] {latdimNDCG.txt};
\legend{Yelp};
\end{axis}
\end{tikzpicture}

\begin{tikzpicture}
\begin{axis}[
    xlabel={\# of GNN Layers $L$},
    ylabel={HR@10},
    xmin=-0.1,xmax=3.1,
    ymin=0.72, ymax=0.84,
    legend columns=1,
    legend cell align=right,
    grid=both,
    every axis plot/.append style={ultra thick},
    every tick label/.append style={scale=1.2},
    label style={scale=2},
    legend style={
        nodes={scale=1.5, transform shape},
        legend image post style={scale=1.5},
        },
    legend style={at={(1,0)},anchor=south east},
    every axis plot post/.append style={
        every mark/.append style={scale=2.2}
    }
]
\addplot[color={rgb:red,4}, mark=o, dashed, mark options={solid}]
table[x=para, y=Yelp] {gnnLayersHR.txt};
\legend{Yelp};
\end{axis}
\end{tikzpicture}

\begin{tikzpicture}
\begin{axis}[
    xlabel={\# of GNN Layers $L$},
    ylabel={NDCG@10},
    xmin=-0.1,xmax=3.1,
    ymin=0.47, ymax=0.6,
    legend columns=1,
    legend cell align=right,
    grid=both,
    every axis plot/.append style={ultra thick},
    every tick label/.append style={scale=1.2},
    label style={scale=2},
    legend style={
        nodes={scale=1.5, transform shape},
        legend image post style={scale=1.5},
        },
    legend style={at={(1,0)},anchor=south east},
    every axis plot post/.append style={
        every mark/.append style={scale=2.2}
    }
]
\addplot[color={rgb:red,4}, mark=o, dashed, mark options={solid}]
table[x=para, y=Yelp] {gnnLayersNDCG.txt};
\legend{Yelp};
\end{axis}
\end{tikzpicture}

\begin{tikzpicture}
\begin{axis}[
    xlabel={\# of Memory Units $|\mathcal{M}|$},
    ylabel={HR@10},
    xmin=1,xmax=17,
    ymin=0.82, ymax=0.84,
    legend columns=1,
    legend cell align=right,
    grid=both,
    every axis plot/.append style={ultra thick},
    every tick label/.append style={scale=1.2},
    label style={scale=2},
    legend style={
        nodes={scale=1.5, transform shape},
        legend image post style={scale=1.5},
        },
    legend style={at={(1,0)},anchor=south east},
    every axis plot post/.append style={
        every mark/.append style={scale=2.2}
    }
]
\addplot[color={rgb:red,4}, mark=o, dashed, mark options={solid}]
table[x=para, y=Yelp] {memoryUnitHR.txt};
\legend{Yelp};
\end{axis}
\end{tikzpicture}

\begin{tikzpicture}
\begin{axis}[
    xlabel={\# of Memory Units $|\mathcal{M}|$},
    ylabel={NDCG@10},
    xmin=1,xmax=17,
    ymin=0.57, ymax=0.59,
    legend columns=1,
    legend cell align=right,
    grid=both,
    every axis plot/.append style={ultra thick},
    every tick label/.append style={scale=1.2},
    label style={scale=2},
    legend style={
        nodes={scale=1.5, transform shape},
        legend image post style={scale=1.5},
        },
    legend style={at={(1,0)},anchor=south east},
    every axis plot post/.append style={
        every mark/.append style={scale=2.2}
    }
]
\addplot[color={rgb:red,4}, mark=o, dashed, mark options={solid}]
table[x=para, y=Yelp] {memoryUnitNDCG.txt};
\legend{Yelp};
\end{axis}
\end{tikzpicture}
    \end{adjustbox}
    
    \caption{Hyper-parameter study on important parametric configurations of \model\ (\ie~the hidden state dimensionality $d$, the number of graph neural iterations $L$, and the number of memory units $|\mathcal{M}|$), in terms of HR@10 and NDCG@10, on Ciao, Epinions, and Yelp datasets.}
    \label{fig:hyperparam}
\end{figure*}
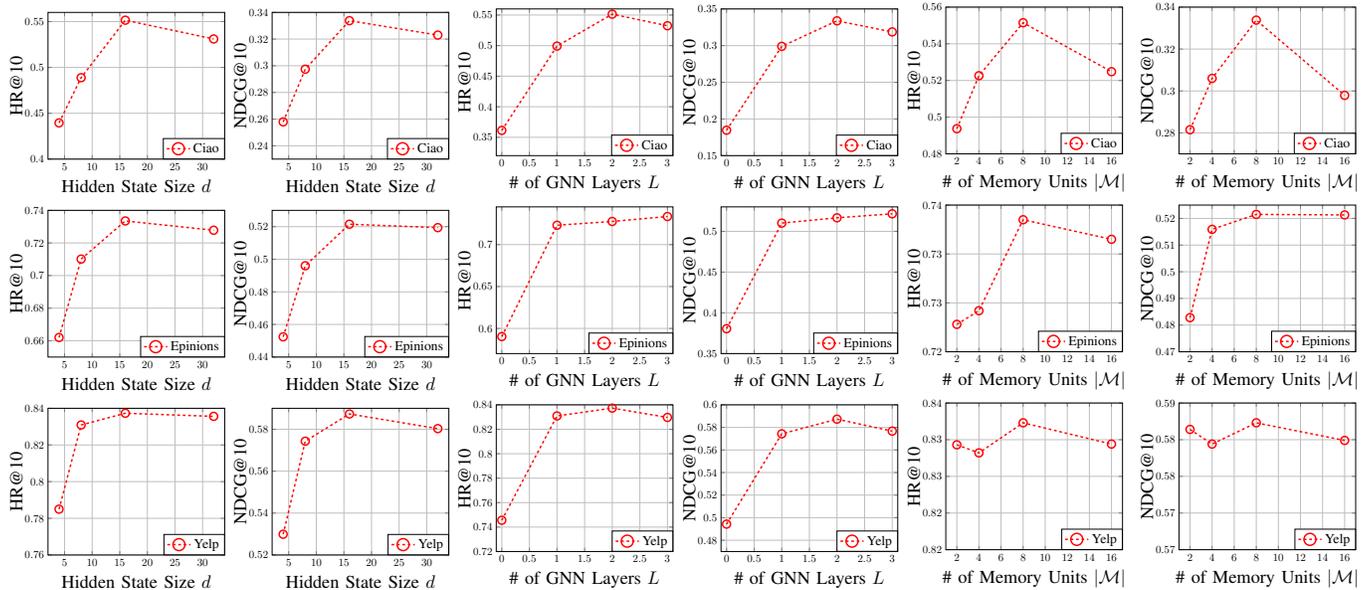

This section studies how the hyperparameter settings affect the performance, by exploring the influences of hidden state dimension size $d$, graph layer numbers $L$, and latent memory unit numbers $\mathcal{M}$. To save space, we present the evaluation results (with different value scales) on different datasets in a unified figure. The y-axis of Figure~\ref{fig:hyperparam} indicates the performance degradation ratio as compared to the best accuracy.\\\vspace{-0.1in}

\noindent \textbf{Hidden State Size $d$}. We plot the performance curve of our \model\ by varying $d$ in range \{$2^2$, $2^3$, $2^4$, $2^5$\}. With the incorporation of semantic relatedness from both user and item domains, we can notice that the embedding dimensionality of 16 is sufficient to bring good performance to our \model\ framework. In addition, the model suffers from the performance degradation with the larger embedding dimensionality. In summary, the strong performance of our model with smaller hidden state dimensionality is beneficial for practical recommender systems, in which the model computational cost is directly influenced by the embedding dimension.\\\vspace{-0.1in}

\noindent \textbf{Graph Layer Numbers $L$}. We perform the context-aware message passing on the collaborative heterogeneous graph $\mathcal{G}$. Now, we investigate the model performance through stacking more graph layers ($1\leq L \leq 3$). We can observe that propagating embeddings across two-hop neighboring nodes brings performance improvement. We also compare our method with the non-propagation variant ($L=0$). The results on three datasets show the effectiveness of encoding high-order relationships between users and items. We also see that by stacking more graph layers, the performance will slightly drop due to the over-smoothing problem of graph neural networks.\\\vspace{-0.1in}

\noindent \textbf{Memory Unit Numbers $|\mathcal{M}|$}. In our experiments, the number of memory units $\mathcal{M}$ is searched from \{$2^1$, $2^2$, $2^3$, $2^4$\}. We can observe that the best performance can be achieved with 8 memory units for encoding relation heterogeneity. Our method leverages the multi-dimensional representation space to capture the semantics of heterogeneous connections by disentangling the implicit factors. 

\subsection{Model Efficiency Study (RQ6)}

We evaluate the model efficiency from two aspects, \ie, the computational cost measured by running time as well as the convergence \wrt\ epochs. Two state-of-the-art baselines (\ie, DGCF and HGT) are experimentally compared.


\subsubsection{Running Time per Epoch}
As shown in Table~\ref{tab:trainTime}, we can observe that our \model\ can achieve better efficiency compared with the disentangled recommender system (\ie, DGCF) and heterogeneous GNN-based method (\ie, HGT). To be specific, HGT employs the transformer-like multi-head dot-product attention mechanism, which is time-consuming. The larger graph size will result in the increased training time of HGT enormously. In addition, for the method DGCF, the cost associated with the design of recursive routing mechanism is huge, which leads to the heavy computational burden for performing the propagation among multiple user embeddings. \\\vspace{-0.12in}


\begin{table}[]
    \centering
    \caption{Running time (seconds) in one epoch for different models.}
    \label{tab:trainTime}
    \begin{tabular}{c|ccc|ccc}
        \hline
         \multirow{2}{*}{Model} & \multicolumn{3}{c|}{Traning} & \multicolumn{3}{c}{Testing}  \\
         \cline{2-7}
         & Ciao & Epinions & Yelp & Ciao & Epinions & Yelp\\
         \hline
         \hline
         DGCF & 2.56 & 63.52 & 81.15 & 0.87 & 14.77 & 66.57\\
         \hline
         HGT & 4.21 & 525.94 & 728.23 & 0.77 & 13.55 & 79.74\\
         \hline
         \emph{\model} & 2.47 & 31.60 & 39.50 & 0.71 &7.39 & 25.57\\
         \hline
    \end{tabular}
\end{table}

\subsubsection{Performance \textit{v.s.} Number of Epochs}
\begin{figure}
    \centering
    \includegraphics[width=0.32\columnwidth]{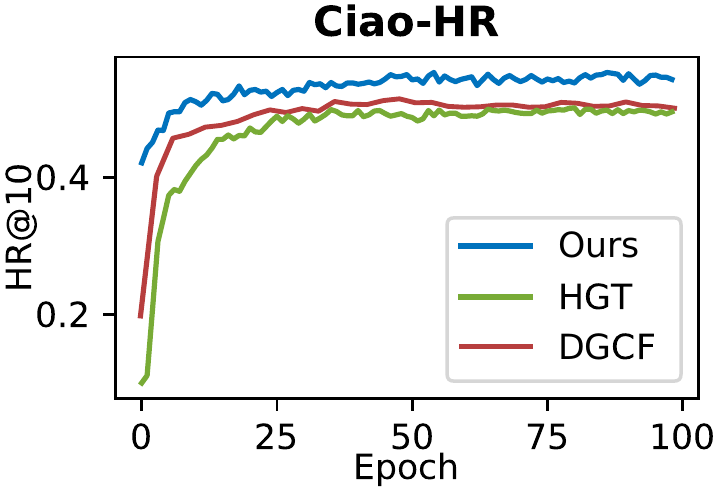}
    \includegraphics[width=0.32\columnwidth]{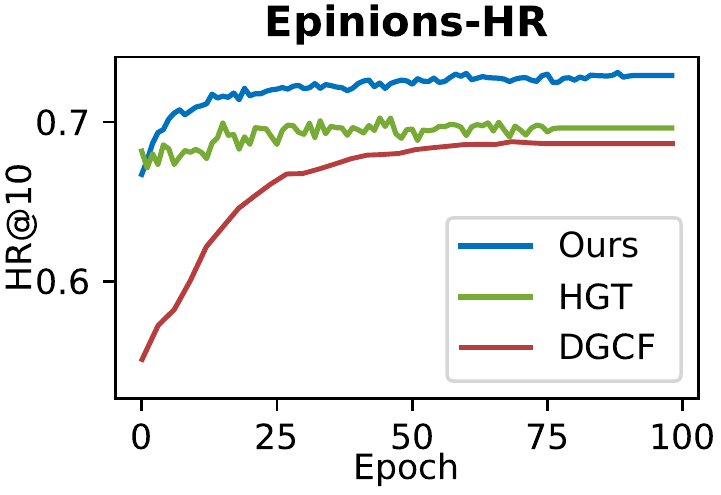}
    \includegraphics[width=0.32\columnwidth]{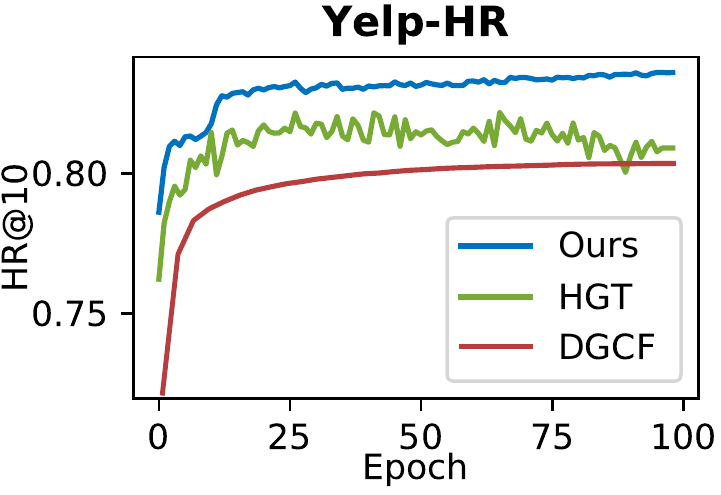}\\\vspace{0.05in}
    \includegraphics[width=0.32\columnwidth]{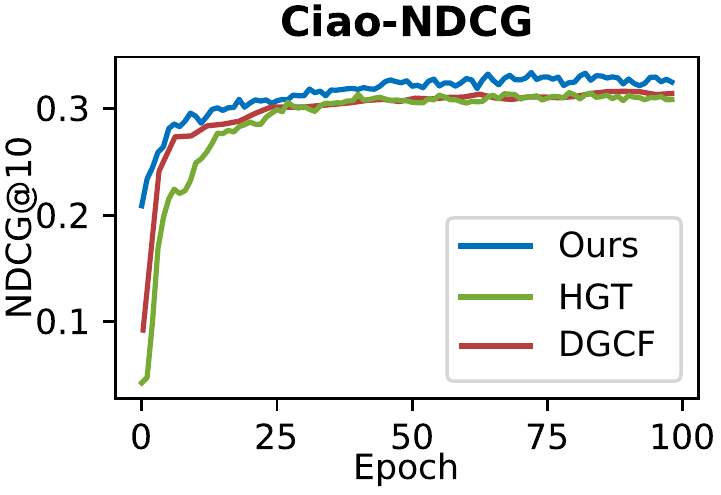}
    \includegraphics[width=0.32\columnwidth]{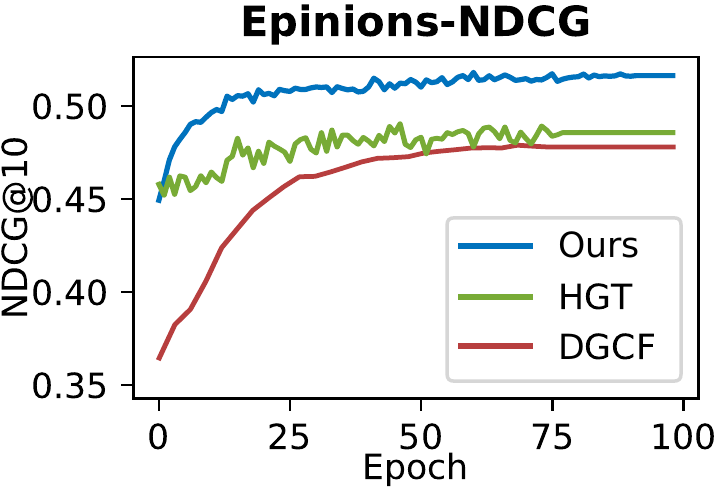}
    \includegraphics[width=0.32\columnwidth]{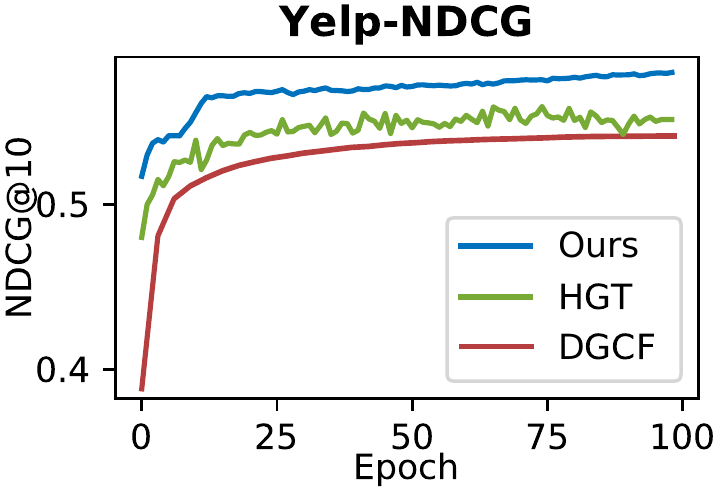}
    \caption{Tested model performances for different methods, \wrt\ the number of training epochs, in terms of {HR@10} and NDCG@10 on the three datasets.}
    \label{fig:convergence}
\end{figure}

We evaluate the model performance in terms of HR@10 and NDCG@10 after each training epoch, which shows the performance improvement with the parameter optimization process. We show the results in Figure~\ref{fig:convergence}, from which we have the following observations: i) \model\ achieves best performance in all epochs compared to HGT and DGCF. This indicates the effectiveness of our model optimization for parameter inference. The architecture of \model\ not only produces better recommendation accuracy, but also are easier to be optimized. We ascribe this to mapping edge relations into multiple latent spaces with the consideration of relation heterogeneity. ii) Compared to DGCF, the performance of HGT increases faster in the early epochs. This implies the advantages of heterogeneous graph transformer over vanilla GCNs in handling heterogeneous graph structures. The reason for such superiority may be the modeling of heterogeneous semantics and the utilization of multi-head dot-product attention technique. \\\vspace{-0.12in}


\begin{figure}[t]
    \centering
    \subfigure[][KGAT]{
        \centering
        \includegraphics[width=0.295\columnwidth]{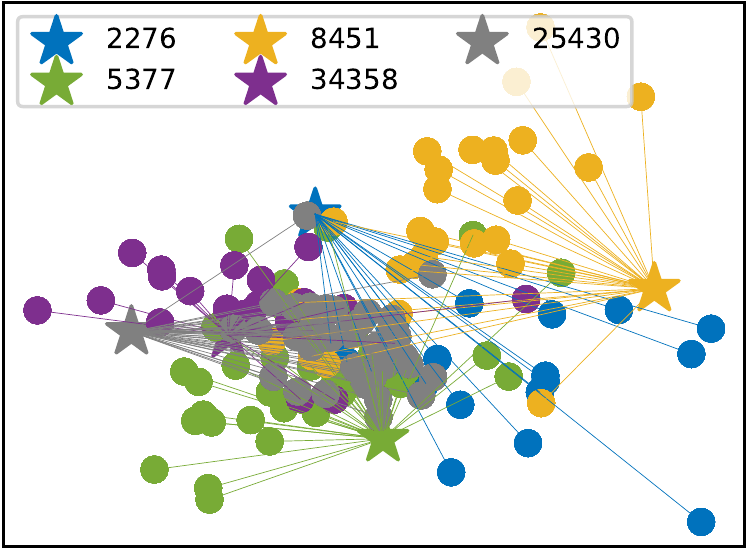}
        \label{hgmn_visualization}
    }
    \subfigure[][HAN]{
        \centering
        \includegraphics[width=0.295\columnwidth]{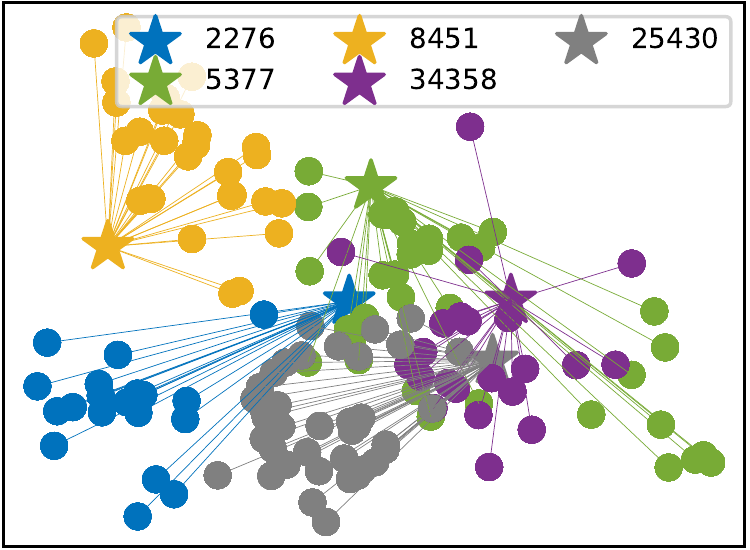}
        \label{ngcf_visualization}
    }
    \subfigure[][\model]{
        \centering
        \includegraphics[width=0.295\columnwidth]{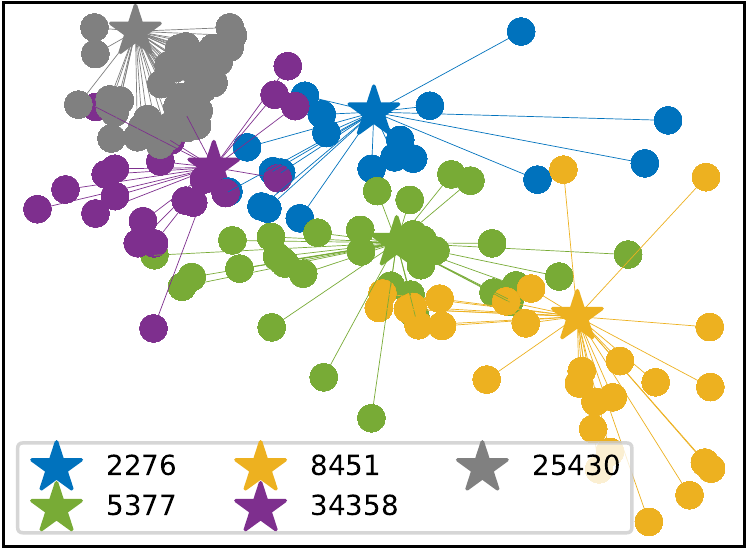}
        \label{ngcf_visualization}
    }
    \vspace{-0.1in}
    \caption{Visualized embeddings for users (stars) and their interacted item (circles), learned by KGAT, HAN and \model. Best view in color.}
    \label{fig:visualization}
    \vspace{-0.1in}
\end{figure}

\subsection{Case Study (RQ7)}
In this section, we conduct case study from two aspects, to investigate the representation ability of \model\ for encoding heterogeneous semantic with disentangled representations. 

\subsubsection{Embedding Visualization}
In this part, we project the learned user/item feature vectors into low-dimensional latent space using t-SNE~\cite{van2008visualizing}. The embedding visualization results are presented in Figure~\ref{fig:visualization}. Different colors in this figure show different users (stars) and their interacted items (circles). The user embeddings encoded by KGAT, HAN and our \model\ method are shown in Figure~\ref{fig:visualization} (a)-(c), respectively. From the results, we summarize the following observations:


\begin{itemize}[leftmargin=*]

\item The better node separation phenomenon of heterogeneous graph neural network (HAN) as compared to knowledge-aware recommender (KGAT), indicating that the exploration of item-wise relationships with social effects will improve the discriminative ability of user preference embeddings. \\\vspace{-0.12in}

\item Given interaction patterns over different items of those sampled users, it is obvious that our \model\ separates users better than other baselines. Also, the item nodes with same colors are commonly distributed around the related user in the center. The low-dimensional representation clearly shows the better ability of \model\ in preserving the user-item relatedness. This superior representation ability should be attributed to the joint modeling of user-user and item-item dependencies, as well as the multi-channel disentangled projection via disentangled graph neural networks.

\end{itemize}

\subsubsection{Memory Attention Visualization}
\begin{figure}
    \centering
    \includegraphics[width=0.95\columnwidth]{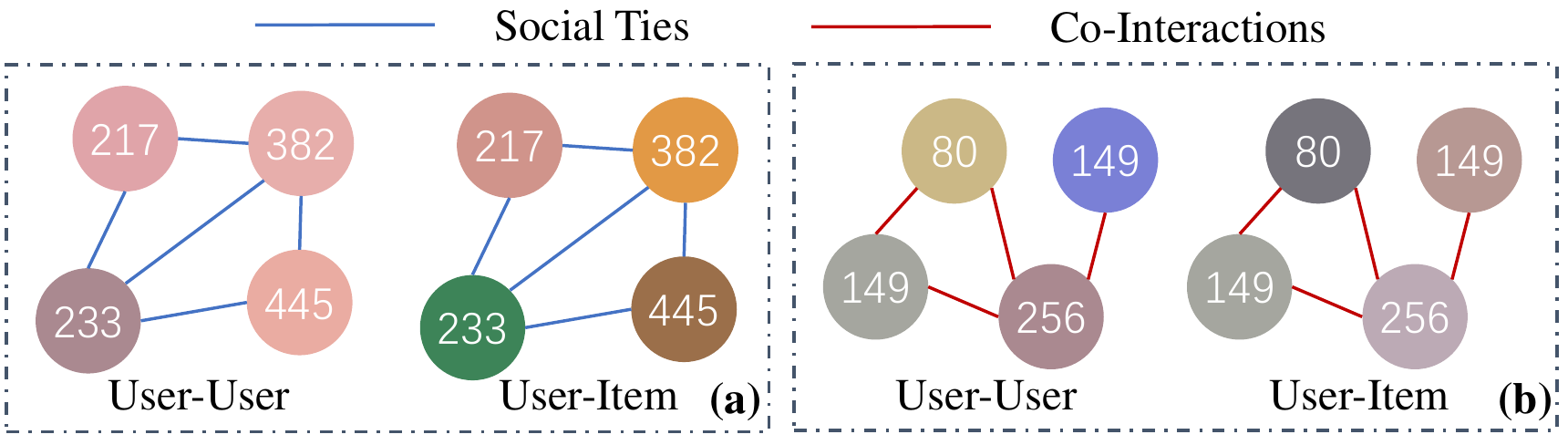}
    \vspace{-0.05in}
    \caption{Visualized users' memory attention vectors learned by our memory-augmented graph encoder, with heterogeneous semantic preserved.}
    \label{fig:memovisual}
    \vspace{-0.1in}
\end{figure}

Next, we visualize the learned attention weights over different memory units for different heterogeneous relations, to validate if the edge-type-specific memory attention of users reflect the corresponding user-wise relations. Specifically, we train a neural network to map the attention vectors $[\eta(\textbf{H}^{(L)}[u_i], 1), ..., \eta(\textbf{H}^{(L)}[u_i], m),..., \eta(\textbf{H}^{(L)}[u_i], M)]$ into three-dimensional vectors corresponding to RGB color values. The mapping function is trained with self-discrimination task, so that the memory attention weights can be visualized with the original user-memory connections preserved.
We randomly pick two closely-connected user-wise subgraphs from Ciao data, one of which contains only social ties, and another one contains only co-interaction relations. The subgraphs are shown in Figure~\ref{fig:memovisual}, where circles represents users with their color denoting the learned memory attention weights. The lines denote the heterogeneous user-wise connections. The embedding visualization is conducted for both the user-user relations, and the user-item relations.

From the results, we can clearly observe that users connected by social ties have close user-user memory-unit weights, while they usually have very different user-item memory weights. This pattern also holds for the co-interaction subgraph, where users having co-interactions have similar user-item memory weights, and have very different user-user memory weights. This observation between relation-specific memory weights and relation-specific graph structures, validates the capability of the memory-augmented heterogeneity encoder of our \model\ in capturing node/edge-type-specific semantics by disentangling diverse relational factors.

\section{Conclusion}
\label{sec:conclusion}

In this paper, we develop a novel Disentangled Graph Neural Network (\model) to improve social recommendation with the disentanglement over heterogeneous relations from both user and item domains. Specifically, by integrating the latent memory units into the graph neural network, \model\ could automatically captures heterogeneous structural dependency with disentangled representations. The model differentiates the relation-aware information propagation and aggregation among the correlated users and items over the collaborative heterogeneous graph. Experimental results show that \model\ can lead to significantly better performance compared to state-of-the-arts on several real-world recommendation datasets. In the future, we would like to endow our \model\ with the power of cross-domain knowledge transfer, so as to further improve the model performance for cold-start recommendations. In addition, another potential direction is to explore the heterogeneous relational data under a pre-trained framework to augment the side knowledge learning.


\bibliographystyle{abbrv}
\balance
\bibliography{refs_full} 

\end{document}